\documentclass[a4paper,11pt]{article}
\pdfoutput=1 

\def\Mh{M_{h}}
\def\Mmin{M_{(h,\rm min)}}
\def\Nion{N_{\rm ion}}
\def\Rmfp{R_{\rm mfp}}
\newcommand{\HI}{\rm H{\scriptsize I}}

\def\impc{{\rm Mpc}^{-1}}

\def\xb{\bar{x}_{\rm H{\tiny I}}}
\def\Tb{T_{\rm b}}

\def\kk{{\vec{k}}}
\def\xx{{\vec{x}}}
\def\Nk{N_{k_{i}}}
\def\cov{\boldsymbol{\sigma}^2}
\usepackage{jcappub} 
\usepackage{psfrag,color,epsfig}
\usepackage{graphicx,graphics}	
\usepackage{upgreek}
\usepackage{amsmath}	
\usepackage{amssymb}	
\usepackage{multicol}	
\usepackage{bm}		
\usepackage{pdflscape}	
\usepackage{multirow}
\usepackage{tabularx}  
\usepackage{physics}
\usepackage{xspace}

\usepackage[T1]{fontenc} 
\usepackage{ae,aecompl}
\usepackage{newtxmath}
\usepackage{ulem}

\title{Improving constraints on the reionization parameters using 21-cm bispectrum}

\author[a,1]{Himanshu Tiwari,\note{Corresponding author}}
\author[b,c]{Abinash Kumar Shaw,}
\author[a,d]{Suman Majumdar,}
\author[a]{Mohd Kamran,}
\author[a,c]{Madhurima Choudhury}

\affiliation[a]{Department of Astronomy, Astrophysics and Space Engineering, Indian Institute of Technology Indore, Indore, India -- 453552.}
\affiliation[b]{Department of Physics, Indian Institute of Technology Kharagpur, Kharagpur, India -- 721302.}
\affiliation[c]{Astrophysics Research Centre, Open University of Israel, Ra'anana 4353701, Israel.}
\affiliation[d]{Department of Physics, Blackett Laboratory, Imperial College, London SW7 2AZ, U. K.}

\emailAdd{himanshuhimang@gmail.com}
\emailAdd{abinashkumarshaw@gmail.com}

\abstract{Radio interferometric experiments aim to constrain the reionization model parameters by measuring the 21-cm signal statistics, primarily the power spectrum. However the Epoch of Reionization (EoR) 21-cm signal is highly non-Gaussian, and this non-Gaussianity encodes important information about this era. The bispectrum is the lowest order statistic able to capture this inherent non-Gaussianity. Here we are the first to demonstrate that bispectra for large and intermediate length scales and for all unique $k$-triangle shapes provide tighter constraints on the EoR parameters compared to the power spectrum or the bispectra for a limited number of shapes of $k$-triangles. We use the Bayesian inference technique to constrain EoR parameters. We have also developed an Artificial Neural Network (ANN) based emulator for the EoR 21-cm power spectrum and bispectrum which we use to remarkably speed up our parameter inference pipeline. Here we have considered the sample variance and the system noise uncertainties corresponding to $1000$ hrs of SKA-Low observations for estimating errors in the signal statistics. We find that using all unique $k$-triangle bispectra improves the constraints on parameters by a factor of $2-4$ (depending on the stage of reionization) over the constraints that are obtained using power spectrum alone.}

\arxivnumber{2108.07279}

\begin{document}
\maketitle
\flushbottom

\section{Introduction}
\label{1}
In the history of our Universe, there was a period when the first luminous sources, i.e. the first stars, galaxies, quasars etc. were formed, and they produced huge amount of Ultraviolet (UV) photons. The UV photons thus produced gradually ionized the neutral hydrogen (\HI) in the Inter-Galactic Medium (IGM). This period is popularly known as the Epoch of Reionization (EoR) \cite{choudhury2003,Bharadwaj_2004,Furlanetto_2006,Choudhury_2009,Pritchard_2012}. The EoR is a least understood period in the history of our Universe. Various indirect observations such as the CMBR brightness temperature fluctuations \cite{komatsu11, planck16}, quasar absorption spectra at high redshifts \citep{fan03, goto11, becker15, barnett17} etc. together suggest that the reionization was an extended process and lasted until the redshift $z \sim 6$ (see e.g. \cite{Robertson_2015,Mitra_2015,Mitra_2018,Dai_2019} etc.).

The observations of the redshifted 21-cm signal, arising due to the spin-flip transition of the electron-proton system from the parallel to anti-parallel states in the rest frame of \HI~ atom, is a most direct and promising way of mapping the \HI~ distribution in the IGM through cosmic time \citep{Sunyaev_1972,Hogan_1979}. Also, the 21-cm signal intrinsically carries the information about the underlying dark matter distribution and the properties of the ionizing source. Therefore it should be able to track the reionization history, i.e. the variation in average ionization state of the IGM with redshift during the EoR.

Motivated by this, a number of first generation radio interferometers such as the GMRT\footnote{\url{http://www.gmrt.ncra.tifr.res.in}} \citep{Paciga13}, 
LOFAR\footnote{\url{http://www.lofar.org/}} \citep{Mertens_2020},
MWA\footnote{\url{http://www.mwatelescope.org/}} \citep{barry19}, 
PAPER\footnote{\url{http://eor.berkeley.edu/}} \citep{kolopanis19} and 
HERA\footnote{\url{https://reionization.org/}} \citep{HERA_2021} have dedicated a considerable amount of their observing time to achieve the first statistical detection of the 21-cm signal from the EoR. However, the detection of this signal has not been possible yet due to various observational obstacles such as the presence of several orders of magnitude stronger foreground emission \citep{dimatteo02,jelic08, Ali_2008, Ghosh_2012} in the same frequency range, system noise \citep{morales05,mcquinn06} etc. Further, the direct imaging of the \HI~ distribution is difficult to achieve using these first generation experiments due to their low sensitivity. These interferometers are instead expected to probe the characteristics of the 21-cm signal fluctuations through a popular Fourier statistics i.e. the power spectrum \citep{Pober_2014, Patil_2017}. Once detected the power spectrum is expected to carry several crucial features of the signal \citep{Bharadwaj_2005, barkana05,lidz08,Choudhury_2009_rev,jensen13,Majumdar_2013, Majumdar_2015,Shaw_2019}. However these existing radio interferometers have measured only few weak upper limits on the 21-cm signal power spectrum to date \citep{Paciga13,barry19,kolopanis19,li19,Mertens_2020,trott20,Yoshiura_2021,HERA_2021}. It is expected that the upcoming humongous SKA\footnote{\url{http://www.skatelescope.org/}} \citep{koopmans2015cosmic, mellema15, Shaw_2019} telescope will have enough sensitivity to make high resolution images of the \HI~ distribution during the EoR, and thus it will be able to probe finer details in the EoR 21-cm signal fluctuations than what is possible using just the power spectrum.

The power spectrum in principle can fully describe the EoR 21-cm signal if it were a Gaussian random field. The non-random distribution of the ionizing sources in the IGM and the gradual growth of the ionized regions surrounding them make the EoR 21-cm signal highly non-Gaussian (see e.g. \cite{Bharadwaj_2004,mellema06,watkinson15, shimabukuro15a,Mondal_2015, Majumdar_2018, Majumdar20, hutter19, Shaw_2020} etc). This inherent non-Gaussianity in the EoR 21-cm signal evolves as the reionization progresses. The power spectrum is unable to capture this time evolving non-Gaussianity. In order to capture the non-Gaussianity present in the EoR 21-cm signal, one requires the higher-order statistics such as the bispectrum \citep{Bharadwaj_2004,Yoshiura_2015, Shimabukuro_2017b,watkinson17,Watkinson_2019,watkinson_2021,Majumdar_2018,Majumdar20,hutter19,Saxena_2020,kamran21, kamran21b, Ma_2021, Mondal_2021}, trispectrum \citep{lewis11,Mondal_2015,Shaw_2019} etc. The bispectrum, the Fourier transform of the three point correlation function, is the lowest order statistic which can capture the non-Gaussian features present in the EoR 21-cm signal at different length-scales. The power spectrum of a field is always positive. On the other hand, the bispectrum can be both positive and negative depending of the nature of the fluctuations in the field. Likewise, the bispectrum carries extra information compared to the power spectrum in case of a non-Gaussian field. The non-Gaussianity in EoR 21-cm signal is arising due to the physical processes that goes on in the IGM, which are directly connected with the astrophysical parameters that govern these physical processes. Therefore, the bispectrum is expected to be highly sensitive to these astrophysical parameters compared to the power spectrum and thus would possibly be able to provide a better constraint on the IGM parameters.

In this work, we use the Bayesian inference technique to estimate the astrophysical model parameters of the EoR using the 21-cm bispectrum. There are several attempts that has been made earlier in this regard. All of the earlier studies are somewhat limited in this regard as they consider only a few specific shapes of triangles in the Fourier space (aka $k$-triangles) such as equilateral \citep{Yoshiura_2015,Shimabukuro_2017b} and isosceles \citep{watkinson_2021} $k$-triangles for their analysis. However it is quite obvious that one can, in principle, construct a huge variety of triangles (both in size and shape) in the $k$-space for which the bispectrum can be estimated. Bispectra estimated for different shapes and sizes of triangles are expected to contain different information about the non-Gaussianity in the signal. Therefore it is apparent to identify all unique triangles in the $k$-space which will have unique information about the signal. Using two shape parameters, \cite{Bharadwaj_2020} have provided a way to identify all unique $k$-triangles for bispectra estimation. Our aim in this paper is to use all unique triangle shapes for a range of possibly detectable $k$ modes for estimation of the EoR 21-cm bispectra and investigate if the inclusion of all unique shapes of triangles provides us a better constraints on the EoR model parameters or not.

One of the major impediment for building an efficient parameter estimation pipeline for the EoR 21-cm observations is the computing cost for generating the model observables of the signal for a large set ($\sim 10^5$ or more) of parameter values in a multi-dimensional parameter space. The model signals are generally produced via either semi-numerical (e.g. \cite{Mesinger_2010, Choudhury_2009,Majumdar_2013, Majumdar_2014, Mondal_2015}) or radiative transfer (e.g. \cite{mellema06,thomas09, ghara15a,ghara15b} etc.) simulations. These simulations are the most computationally expensive component of a Bayesian inference process. Additionally, the estimation of higher-order statistics e.g. bispectrum from the simulated signal in a large volume, as well as for a large dynamic range also requires significant amount of computing resources \citep{watkinson17, Majumdar_2018, Majumdar20,kamran21, Mondal_2021, Shaw_2021}.

One approach that has been adopted recently by several groups to circumvent theses obstacles is to use emulators for the EoR 21-cm signal statistics instead of actual simulations. In this context, \cite{Agarwal_2012,Agarwal_2014} have used Artificial Neural Network (ANN) emulation techniques to predict highly accurate non-linear power spectrum. \cite{Kern_2017} have used Gaussian Process (GP) regression based signal power spectrum emulators to constrain the astrophysics and cosmology of the EoR. In a different approach, \cite{ChoudhuryM_2020, ChoudhuryM_2021} have used ANNs to directly predict the EoR parameters from the simulated signal statistics.

In this work we adopt the formalism proposed by \cite{Schmit_2018}. They have developed an ANN based emulator for EoR 21-cm power spectrum using a large training set of simulated signal. They next used this signal power spectrum emulator as their model statistics to constrain EoR parameters via a Bayesian inference process. Extending their approach further, we have developed emulators for both the power spectrum and bispectrum of the EoR 21-cm signal using the ANN. In case of the signal bispectrum emulation we also make sure that it emulates the bispectra for all unique $k$-triangle shapes and for a variety of triangle sizes, therefore ensuring none of the unique features of the signal bispectrum is missed while we use it for EoR parameter estimation via our Bayesian inference pipeline. We use a large database of simulated signal statistics, simulated via a semi-numerical algorithm \texttt{ReionYuga} \citep{Majumdar_2014, Mondal_2015}, to train our emulators. To make sure that this training is done in an efficient and optimal manner the multi-dimensional EoR parameter space is sampled via Latin-Hypercube (LH) method following \citep{Schmit_2018}, and the signal statistics are simulated at these sampled points to generate the training sets for the emulators. Once the signal emulators are trained and tested we used them to estimate the EoR parameters via a Bayesian inference pipeline. In the parameter estimation exercise we have considered the uncertainties due to the sample variance and system noise for instruments like the SKA-low. We next check the performance of signal power spectrum, bispectrum for specific $k$-triangle shapes and bispectra for all unique $k$-triangle shapes in constraining EoR parameters.

The content of this paper is organized in the following manner. In \S\ref{2}, we describe the statistical estimators of EoR 21-cm signal (i.e. 21-cm power spectrum and bispectrum) which we use in our parameter inference pipeline. \S\ref{3} describes the simulations along with the reionization model and its parameters. \S\ref{4} explains our emulation technique in details. We present the results of our entire analysis in \S\ref{5} and summarize our findings in \S\ref{7}.


\section{Statistical Observable}\label{2}
The radio-interferometric observations are expected to measure fluctuations in the brightness temperature distribution $\delta \Tb(\xx)$ of the redshifted 21-cm radiation from EoR (see e.g. \citep{Choudhury_T_2006, Furlanetto_2006, Pritchard_2012} etc. for reviews). Various statistical tools, in real space as well as in Fourier space, are employed to quantify the information encrypted in these fluctuations. This work uses the power spectrum (two-point) and the bispectrum (three-point) statistics to interpret the EoR 21-cm signal in Fourier space.

\emph{Power spectrum:} The Power Spectrum (PS) is the primary statistic that will be measured by the EoR interferometric experiments. Considering any wave number $\kk$, the EoR 21-cm PS can be written as $P(\kk)=V^{-1}\langle \Delta_{\rm b}(\kk) \Delta_{\rm b}(-\kk) \rangle$. Here $V$ is the signal volume, $\langle \cdots \rangle$ denotes the ensemble average and $\Delta_{\rm b}(\kk)$ is the Fourier conjugate of $\delta \Tb(\xx)$. However, we employ a spherically averaged binned power spectrum (SAPS) estimator, which for any $i$-th bin is defined as
\begin{equation}
    \Bar{P}(k_i)=\frac{1}{\Nk} \sum_{\kk_{a}} P(\kk_a)~.
    \label{eq:PS}
\end{equation}

Here $k_i$ is the average Fourier mode of the $i$-th  bin, the summation runs over all the modes $\kk_a$ within the bin and $\Nk$ is the corresponding total number of modes. However instead of eq. (\ref{eq:PS}), we use the dimensionless power spectrum $\Delta^2(k)= k^3 \bar{P}(k)/(2\pi^2)$ in our subsequent discussions. Note that, here we consider logarithmically separated spherical shells in $\kk$ space to be the bins as described in \cite{Mondal_2015}.

\emph{Bispectrum:} The PS, being a two-point statistics, is unable to quantify the non-Gaussianity present in the EoR 21-cm signal. The three-point correlation function or its Fourier conjugate the bispectrum (BS) is the lowest-order statistics which is sensitive to the non-Gaussianity. Various important astrophysical information of the EoR is expected to be encoded in its non-Gaussian features. The EoR 21-cm BS can be defined as $\delta^{\rm K}(\kk_1+\kk_2+\kk_3)\, B(\kk_1,\kk_2,\kk_3)=V^{-1}\langle \Delta_{\rm b}(\kk_1) \Delta_{\rm b}(\kk_2) \Delta_{\rm b}(\kk_3) \rangle$. The Kronecker's delta here ensures the definition to be valid only for $\kk_1+\kk_2+\kk_3=0$ \textit{i.e.} when the three $\kk_a$'s form a closed triangle. Following the idea of binned estimators, here we consider bins in the triangle configuration space. Corresponding to any $i$-th triangle configuration bin, the binned bispectrum estimator can be written as \citep{Majumdar_2018}
\begin{equation}
    \Hat{B}_i(\kk_1,\kk_2,\kk_3) = \frac{1}{N_{\rm tri}\,V} \sum_{(\kk_1+\kk_2+\kk_3=0)\,\in\,i} \Delta_{\rm b}(\kk_1) \Delta_{\rm b}(\kk_2) \Delta_{\rm b}(\kk_3)~,
    \label{eq:binbs}
\end{equation}
where $N_{\rm tri}$ is the total number of triangles within the corresponding bin. An ensemble average of eq. (\ref{eq:binbs}) yields the bin-averaged bispectrum estimates \textit{i.e.} $\Bar{B}_i(\kk_1,\kk_2,\kk_3)=\langle\Hat{B}_i(\kk_1,\kk_2,\kk_3)\rangle$. 

\begin{figure}
    \includegraphics[scale=0.12]{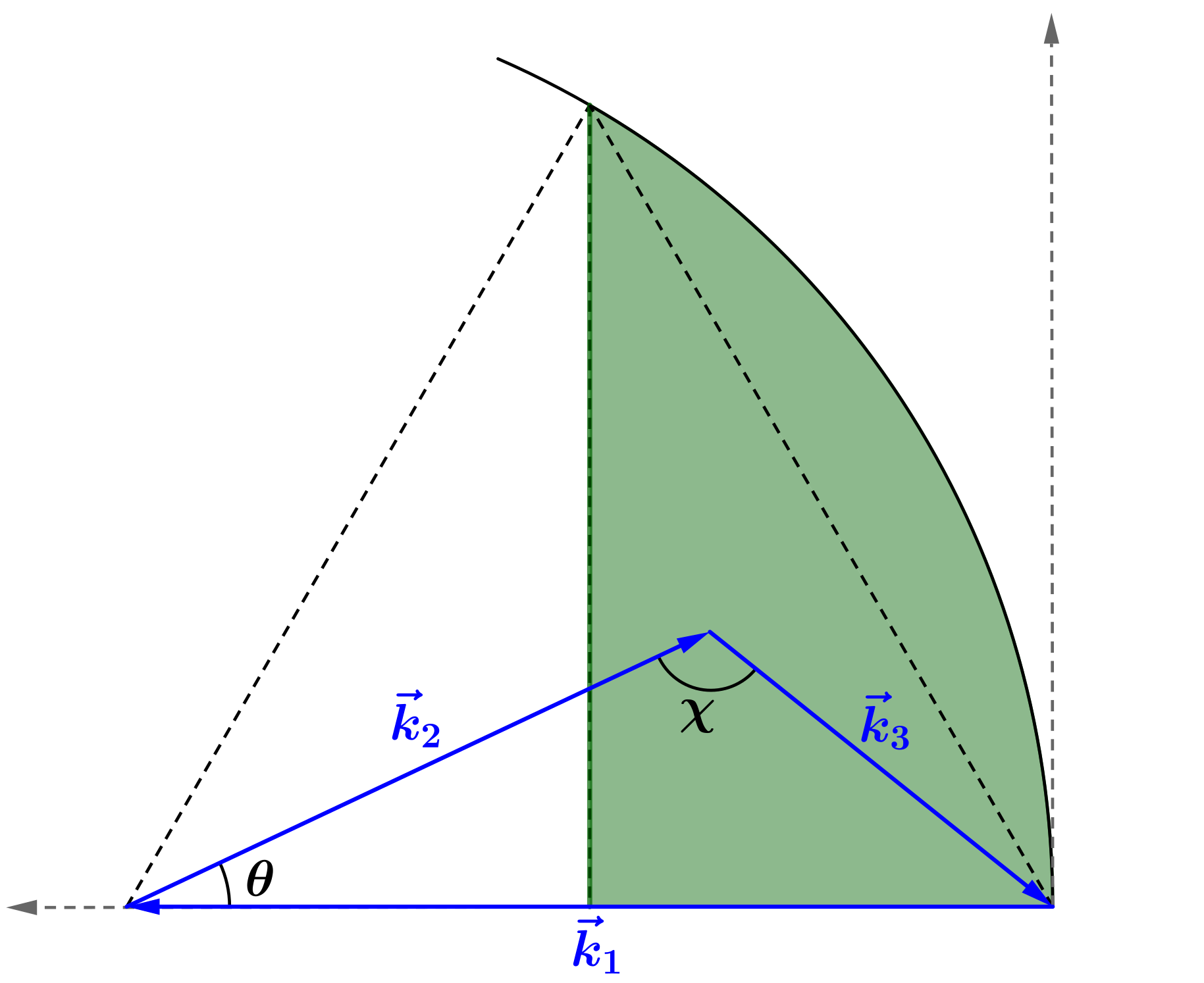}
    \includegraphics[scale=0.6]{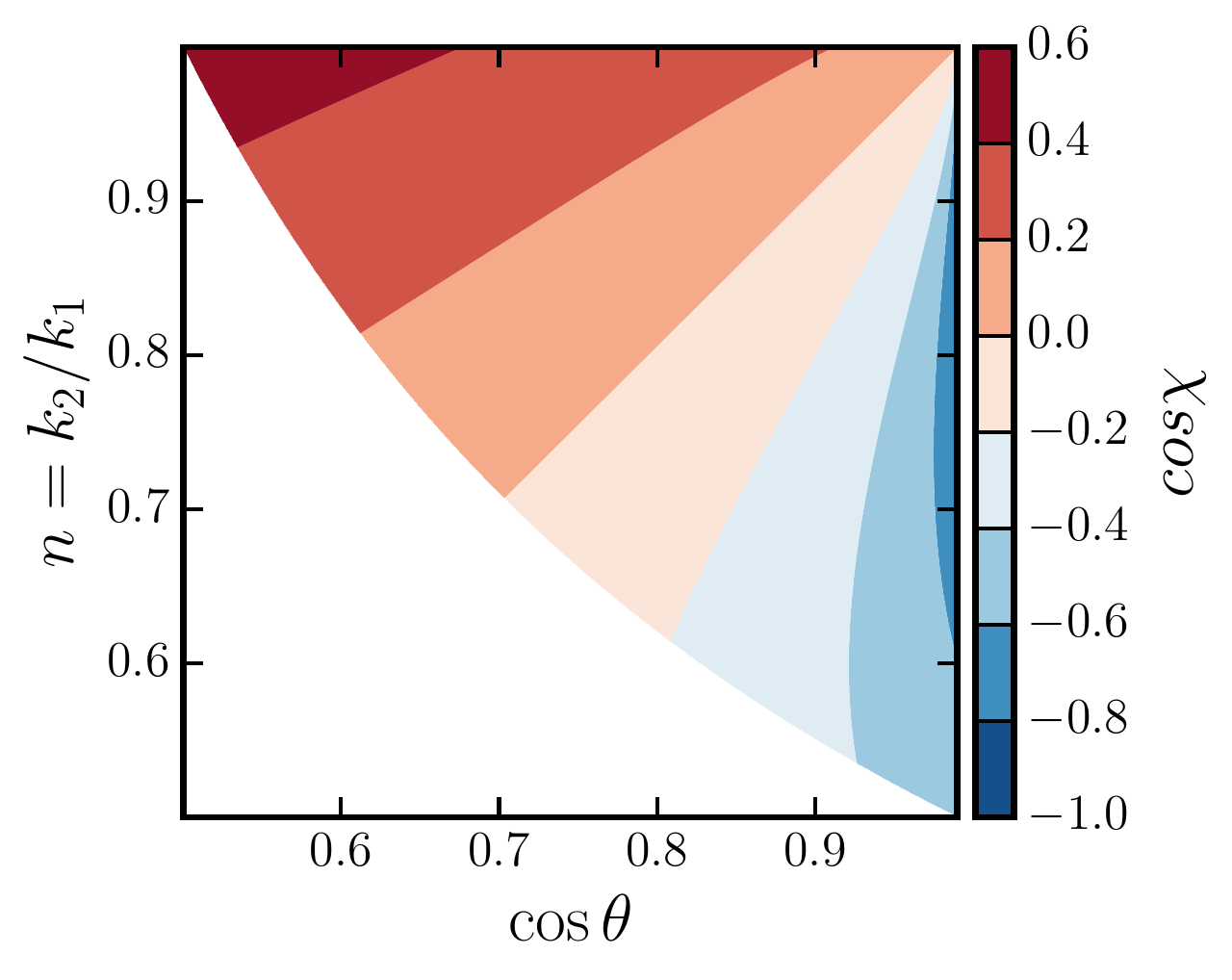}
    \caption{The left panel shows the allowed region (green shade) for $\kk_2$ spanning the triangles of all unique shapes for a fixed $\kk_1$. The right panel shows how the different unique shapes (denoted by $\cos{\upchi}$) of the triangles is distributed over the $n-\cos{\theta}$ plane.}
    \label{fig:unique_triangle_space}
\end{figure}

Considering statistical isotropy of the signal, the EoR 21-cm BS is independent of the orientation of the triangles in space and remains function of only the size and the shape of triangles. We denote the length of the largest side as the size of the triangle, say $k_1$, corresponding to which the shape is defined by
\begin{equation}
\begin{split}
    n &= \frac{k_2}{k_1}\\
    \cos{\theta} &= -\frac{\kk_1 \cdot \kk_2}{k_1\,k_2}~,
\end{split}
\label{eq:shape}
\end{equation}
where $k_a=|\kk_a|$, $n$ is the ratio of the two largest sides and $\theta$ is the angle between $\kk_1$ and $-\kk_2$ vectors (see the top panel of Fig. \ref{fig:unique_triangle_space}). It is convenient to represent the BS using $(k_1,n,\cos{\theta})$ parameterization instead of $(k_1,k_2,k_3)$, and we further use $B(k_1,n,\cos{\theta})$ unless stated otherwise. For a particular size of the triangle (fixed $k_1$), a complete set of triangles with unique shapes can be obtained by following the prescription of \cite{Bharadwaj_2020} which is given as

\begin{equation}
    \begin{split}
        & k_1 \geq k_2 \geq k_3\\
        & 0.5 \leq \cos{\theta} \leq 1.0\\
        & 0.5 \leq n \leq 1.0~.
    \end{split}
    \label{eq:cond}
\end{equation}

Considering top panel of Fig. \ref{fig:unique_triangle_space}, the $\kk_2$ is restricted to the green-shaded region in order to satisfy the above conditions for a particular $\kk_1$. In the bottom panel, we show the variation of $\cos{{\upchi}}$ in $n$--$\cos{\theta}$ plane that account for different unique shapes of triangles (equilateral, isosceles and scalene). We refer the reader to section $2.2$ of \cite{Mazumdar_2020} for a more rigorous discussion on unique triangles. This parametrization for the binned bispectrum is followed in the subsequent discussions unless stated otherwise. The readers are referred to \cite{Majumdar_2018} and \cite{Majumdar20} for a detailed description of the methodology for computing bispectrum.
Here we divide the $\kk$ space (within the Nyquist limit) into $15$ logarithmically separated semi-spherical shells of $k_1$ for which we compute the spherically averaged bin bispectrum (SABS). The mean of the $k_1$ shell represents the average size of the binned triangles in our formalism. We also divide the $n-\cos{\theta}$ plane in regular grids with spacings $\Delta n=0.05$ and $\Delta \cos{\theta}=0.01$ respectively for $n$ and $\cos{\theta}$. Similar to the PS, we use the dimensionless form of the bispectrum $\Delta^3(k_1,n,\cos{\theta})=k_1^6 n^3 B(k_1,n,\cos{\theta})/(2\pi^2)^2$ in our final implementation and parameter inference below.


\section{Reionization Model and Simulation}\label{3}

We consider an \textit{inside-out} model for reionization where the sources ionize the surrounding medium first and then the ionizing radiations leak into the IGM. Our model, which closely follows that in \cite{Choudhury_2009}, is based on two fundamental assumptions -- $(1)$ \HI~ follows the underlying matter density contrast, and $(2)$ the ionizing sources form within the dark matter halos having masses above a certain lower cut-off, say $\Mmin$. The ionizing field is generated using a simple but effective prescription where the amount of UV photons $N_\gamma$ diffusing into the IGM is proportional to the halo mass. Here $\Mmin$ and the proportionality constant, say $\Nion$, are the two model parameters which are related to the astrophysics of source formation. In addition to these, $\Rmfp$ is the third parameter in our simulations which denotes the mean free path of the ionizing photons in the IGM. A brief description of the parameters is as follows.
\begin{itemize}
    \item $\bm{M_{(h,\rm \textbf{min})}}$: Not every halo is eligible to participate in the star formation. $\Mmin$ denotes the lower halo mass limit, and any halo above this floor is expected to collapse ample amount of pristine gas which will cool down to form the first generation of sources. Its value is determined by various possible cooling mechanisms (atomic/molecular) of the accreted gas. Increasing the value of $\Mmin$, by keeping other parameters constant, delays the reionization process and vice-versa.
    
    \item $\bm{N_{\rm \textbf{ion}}}$: Following the aforementioned prescription, the amount of ionizing photons produced can be written as
    \begin{equation}
        N_\gamma(\Mh \geq \Mmin)= \Nion \frac{\Omega_{b}}{\Omega_{m}} \frac{\Mh}{m_p}~,
    \end{equation}
    where $\Nion$ is a dimensionless proportionality constant which encapsulates several degenerate source parameters such as the star formation efficiency ($f_{*}$), the escape fraction of ionizing photons ($f_{\rm esc}$) and the recombination rate $N_{\rm rec}$ etc \citep{Choudhury_2009_rev}. $\Omega_{m}$, $\Omega_{b}$ and $m_p$ are respectively the dark matter density parameter, baryon density parameter and proton mass. A larger value of $\Nion$ implies existence of more efficient sources and a larger $N_\gamma$ that hastens the process of reionization.
    
    \item $\bm{R_{\rm \textbf{mfp}}}$: The ionizing photons get readily absorbed by the local medium. However, they tend to cover some distance depending on the nature of the surrounding medium and the photon production rates of host sources. $\Rmfp$ represents the typical size of the ionized region around the sources which, in principle, can vary for changing source and IGM properties. Increasing the $\Rmfp$ values would result in larger ionized regions around the most efficient sources which ends the reionization at a faster pace.
\end{itemize} 
For a more detailed discussion about theses parameters we refer the interested reader to the Section $2$ of \cite{Shaw_2020}.  

Using the aforementioned model of reionization, we simulate the comoving volumes of EoR 21-cm signal by employing an \textit{excursion-set} formalism \citep{Furlanetto_2004}. Producing the reionization maps involves three major steps here. First, the dark matter density field is generated at desired redshifts using a particle-mesh based \textit{N}-body code{\footnote{Publicly available at \href{https://github.com/rajeshmondal18/N-body}{https://github.com/rajeshmondal18/N-body}}} \citep{Bharadwaj_Nbody}. We have simulated the matter density fields within a comoving box of volume $V=[215~{\rm Mpc}]^3$ at $z=8$, which is the redshift of our interest in this analysis. The spatial grid size in this step is $0.07~{\rm Mpc}$ which yields a mass resolution of $1.09\times10^8~{\rm M}_{\odot}$. In the second step, we locate the halos using a Friends-of-Friend (FoF) halo finder code{\footnote{Publicly available at \href{https://github.com/rajeshmondal18/FoF-Halo-finder}{https://github.com/rajeshmondal18/FoF-Halo-finder}}}. We choose the linking-length between two friends to be $0.2$ times the mean inter-particle separations and set the criterion that a minimum of $10$ dark matter particles can form a dark matter halo \citep{Davis_1985}. This yields the halos with minimum mass $1.09\times 10^9~{\rm M}_{\odot}$ in our simulations. In the last step, we generate the 21-cm maps using our semi-numerical code \texttt{ReionYuga}{\footnote{Publicly available at: \href{https://github.com/rajeshmondal18/ReionYuga}{https://github.com/rajeshmondal18/ReionYuga}}} \citep{Majumdar_2014, Majumdar_2015, Mondal_2015}. The code takes the dark matter and the halo distributions as inputs to simulate reionization maps on a grid which is $8$ times coarser than those in the \textit{N}-body simulations. Finally, we simulated a various models-- corresponding to different reionization histories to develop the emulator described in the following section.


\section{Emulating E\lowercase{O}R 21-cm statistics} \label{4}
\begin{figure}
    \centering
    \includegraphics[scale=0.37]{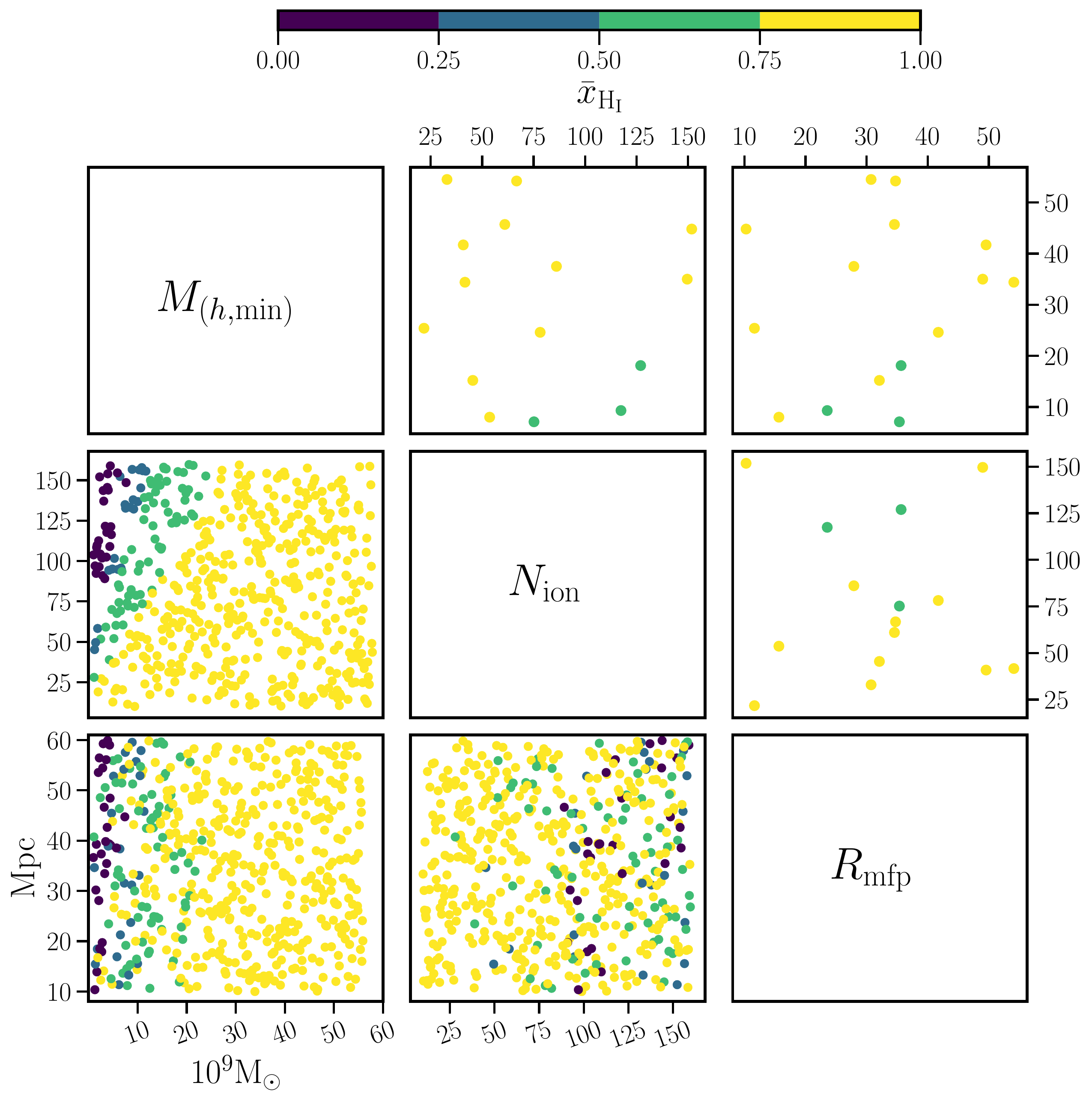}
    \caption{Parameter space defined by Latin-Hypercube sampling; training set (lower off-diagonal panels) and testing set (upper off-diagonal panels). The different colors scheme of the sampled points denote the corresponding value of $\xb$ at $z=8$.}
    \label{fig:parameter_space}
\end{figure}

The central idea behind using artificial neural networks (ANN) is to develop fast and reliable EoR models that can replace computationally expensive and time consuming simulations in the process of Bayesian parameter inference (see \S\ref{5}) from observations. A simplistic approach is to use the regression-based ANN emulator, capable of capturing and reproducing the complicated signatures of EoR in the statistical observables of the target signal, in the context of this paper it is the 21-cm signal. The emulator intakes input features (e.g. reionization parameters, described in the previous section) and returns the 21-cm statistics (21-cm power spectrum/bispectrum). In this way, the emulator behaves much like a simulation-based statistical estimator of the signal with significant speedup. The emulator is trained on a large sample of featured input-output data pairs (i.e. simulated realizations of 21-cm power spectrum/bispectrum [output] for a combination of EoR parameters [input], in this context), a.k.a training set. Once the emulator is sufficiently trained, it can predict the output features for unseen input data (test set) with high accuracy.  However, the sampling and size of the training set affect the emulator's accuracy on the test predictions. Thus, constructing an optimal training set is crucial for the emulator's performance. The Latin-Hypercube (LH) sampling method provides an efficient way to populate the multi-parameter space than a simple gridded sampling approach. The advantage of LH sampling is that no two parameters share the same value in the LH multi-parameter space, and thus provide an all  unique set of parameters (see e.g. \cite{Morris_1995, Heitmann_2009,Heitmann_2014,Heitmann_2016, Urban_2010}). We constructed both our training and test set over the points identified by the LH sampling. We generated $550$ LH samples of parameters ($\Mmin, \Nion, \Rmfp$) (see Fig. \ref{fig:parameter_space}) and simulate the corresponding 21-cm power spectrum and bispectrum using Reion-Yuga. The parameter range is as follows -- 
$\Mmin(10^9 \rm M_\odot) \in [1.0,~ 55.0]$, $\Nion \in [10.0,~ 180.0]$ and $\Rmfp ({\rm Mpc}) \in [20.0,~ 60.0]$. 

For this work we have built two emulators, one each for the EoR 21-cm power spectrum and bispectrum. A similar work \cite{Schmit_2018} showed that a minimum $100$ LH-samples are sufficient to train the emulator for EoR 21-cm power spectrum (PS). However, for 21-cm bispectrum (BS), it would require more training samples to achieve comparable accuracy. Thus, training set of size  $>500$ LH samples are expected to be sufficient for both the statistics.

\subsection{ANN architecture}
\begin{figure}
    \centering
    \includegraphics[scale=.28, angle=270]{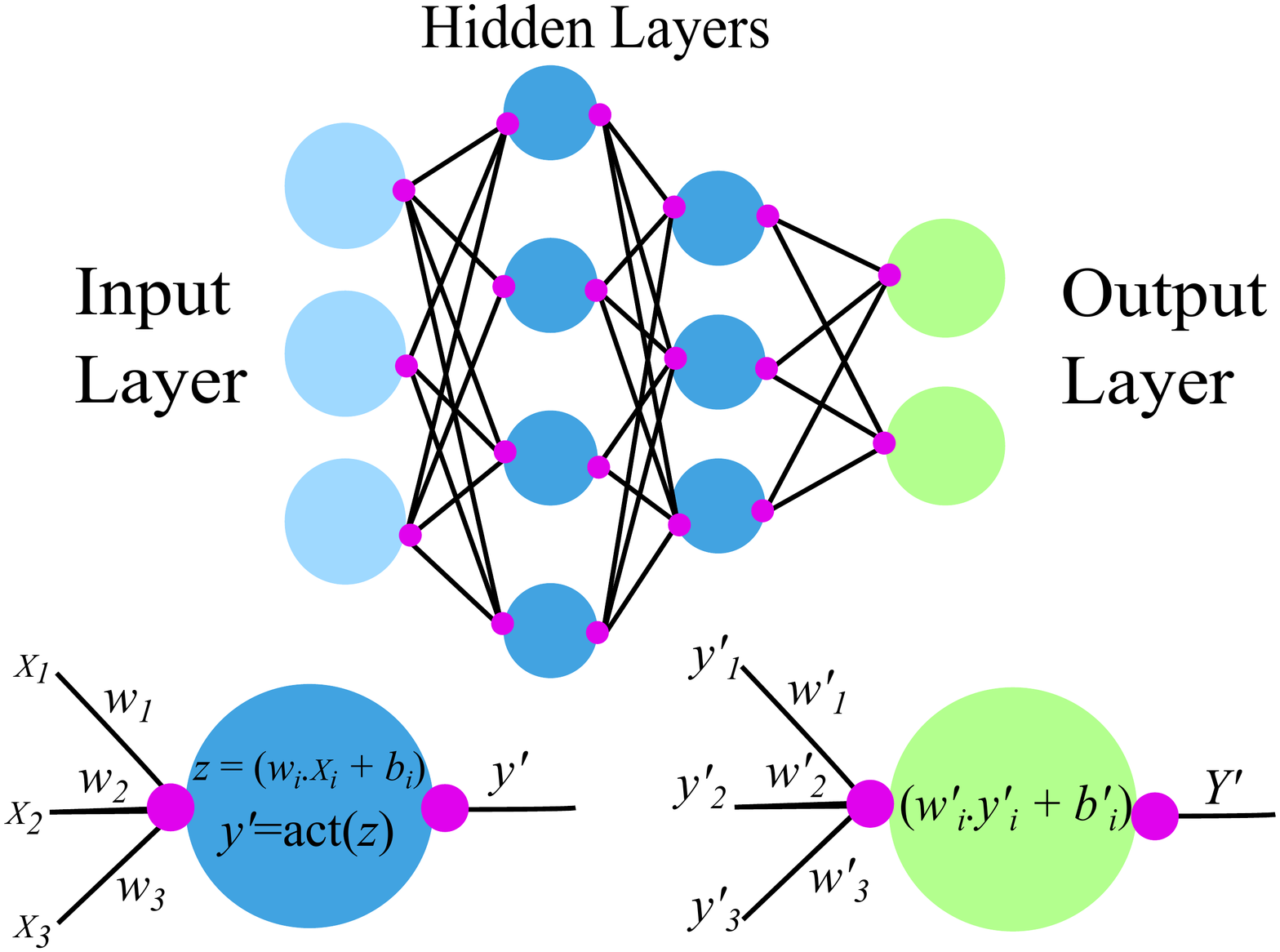}
    \caption{Shows the structure of a typical ANN and its components.}
    \label{fig:ann_structure}
\end{figure}

The fundamental unit of an ANN is an artificial neuron, which primarily is a linear scaling function concatenated with a non-linear activation function. The layers of such interconnected neurons form the neural network (see Fig. \ref{fig:ann_structure}). There are three major components of the neural network; the input layer, the hidden layers and the output layer.
As described previously, the ANN is trained to map the input data/features to the output data. Thus the shape of input and output layers remains equal to the shape of input and output data. We denote input data as $(X)$ which correspond to the reionization parameters $(\Mmin,~\Nion,~\Rmfp)$ and the output data as  $(Y)$ which is the simulated 21-cm statistics (i.e. power spectrum $\Delta^2(k)$ and bispectrum $\Delta^3(k_1,n,\cos{\theta})$). The number of hidden layers and the neuron counts at each hidden layer depends upon the complexity in the training data, and can be adjusted via hyperparameter optimization techniques.

Considering a training data of the form $Y = f(X)$, the prediction from the $j$-th neuron in the $\ell$-th hidden layer can be written as: 
\begin{equation}
    Y^\prime_j = {\rm act}\left(\sum_i^{N_{\ell-1}}{\left[w_{ij}^{\ell}.y^{\prime\ell}_{ij} + b_{ij}^{\ell}\right]}\right)~,
\end{equation}
where the summation is over all the neurons $(N_{\ell-1})$ in the previous layer. Note that $\ell=0$ represents the input layer. Also, $w_{ij}^{\ell}$, $b_{ij}^{\ell}$ and $y^{\prime\ell}_{ij}$ are respectively the weights, bias and predictions linked between $\ell$-th and $(\ell-1)$-th layers. Note that the entire function is encapsulated inside an activation function ${\rm act}(\cdots)$ which indicates that an activation function is applied to every neuron in the network.
The training step aims to tune the weights and bias in such a way that the final ANN predictions ${Y^{\prime}}$ become close to the actual output features ${Y}$. At each training epoch/iteration, the errors between the actual and prediction back-propagates thereby adjusting the weights and bias at each individual node of the network.

In this framework, we optimize our ANN using Mean Squared Error (MSE) loss function which is defined as:
\begin{equation}
    {\rm MSE} = \frac{1}{N_d}\sum_{i=1}^{N_d}\left({Y_i^{\prime}- Y_{i}}\right)^2~,
\end{equation} 
where the summation is over all the $N_d$ number of output data elements. Here we have used {\it Adam optimizer} (see \cite{kingma2017adam} for details) which is an advanced implementation of the popular gradient-descent algorithm. We define the total loss (cost) function of the network as:
\begin{equation}
    {E = \frac{1}{N_{\rm train}}\sum_{n=1}^{N_{\rm train}} E_n(w,b) = \frac{1}{N_{\rm train}}\sum_{n=1}^{N_{\rm train}}\left[\frac{1}{N_d}\sum_{i=1}^{N_d}\left({Y^\prime_{(i,n)}-Y_{(i,n)}}\right)^2\right]}~,
    \label{eq:eq_9}
\end{equation}
where $N_{\rm train}$ denotes the number of training epochs. The weights $(w)$ and the bias $(b)$ at each training epoch are updated in the following way:
\begin{equation}
    {\Delta w_{ij}^\ell = w_{0_{ij}}^\ell -\eta \frac{\partial E}{\partial w_{ij}^\ell} =  w_{0_{ij}}^\ell -\eta \sum_{n=1}^{N_{\rm train}}\frac{\partial E_n}{\partial w_{ij}^\ell}}
\end{equation}
\begin{equation}
    {\Delta b_{ij}^\ell = b_{0_{ij}}^\ell -\eta \frac{\partial E}{\partial b_{ij}^\ell} =  b_{0_{ij}}^\ell -\eta \sum_{n=1}^{N_{\rm train}}\frac{\partial E_n}{\partial b_{ij}^\ell}}~,
\end{equation}
where $w_{0_{ij}}^\ell$ and $b_{0_{ij}}^\ell$ are the initial weights and bias respectively, and $\eta$ is the learning rate. A small value of $\eta$ can consume significant training time whereas a large value of $\eta$ can completely miss the global minima of the loss function. Therefore it is crucial to choose an optimal value for $\eta$ wisely. Here we set $\eta$ and other hyper-parameters using hyperparameter optimization discussed briefly in the following section.

\subsection{Emulating 21-cm Power Spectrum}
\begin{figure}
    \centering
    \includegraphics[scale=0.425]{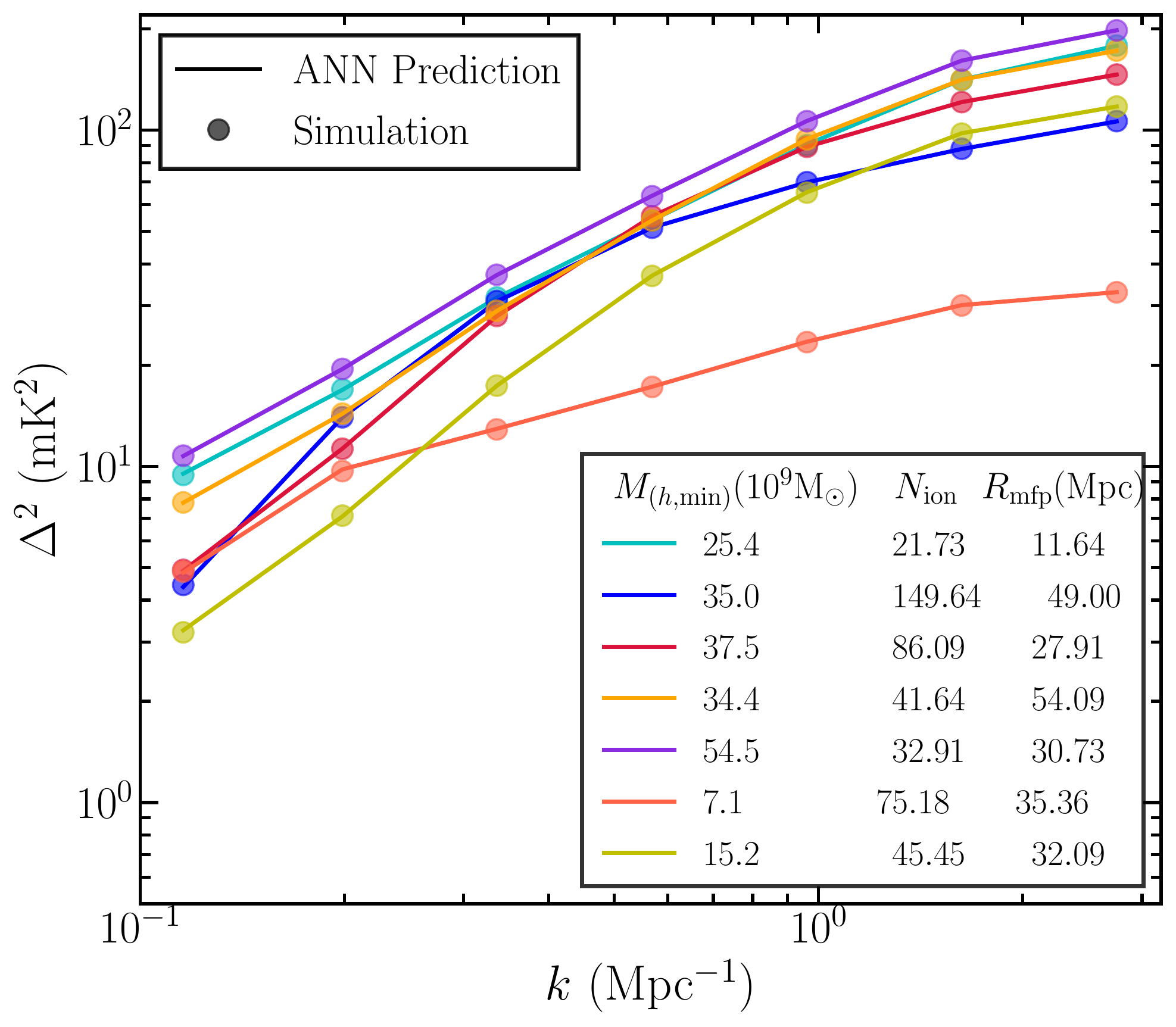}
    \caption[]{Showing the comparison between the emulated power spectrum (solid lines) and the simulation power spectrum (dots) for $7$ test sets. ANN predictions for rest of the test sets is available at our GitHub page\footnote .}
    \label{fig:ann_powerspectrum_emulation}
\end{figure}

We used Python-based deep learning package \texttt{Keras}\footnote{Publicly available at: \href{https://keras.io/}{https://keras.io/}} to develop the ANN emulators for this work. \texttt{Keras} sequential API provides a simple approach to build complex multi-layered ANN structures. We developed the power spectrum emulator by training it with LH sampled data set of reionization parameters $(\Mmin, \Nion, \Rmfp)$ and the simulated 21-cm power spectra (i.e. $\Delta^2(k)$ for $7$ different $k$ modes) for those parameter sets. We have used $90\%$ of $535$ LH samples to train the network and the rest $10\%$ to validate the network's performance during the training, and the remaining $15$ samples were used as the test set.
We used Python-based \texttt{Keras-Tuner} \citep{omalley2019kerastuner} to optimize the hyper-parameters. These are mainly the number of hidden layers, the number of neurons in each hidden layer, loss function, activation function, optimizer, learning rate $\eta$ and batch-size. The best configuration of ANN was found to have two hidden layers with $28$ and $14$ nodes (neurons), respectively. The input and output layers follow the shape of the training data (i.e. $3$ and $7$, correspond to $3$ reionization parameters and $7$ PS values at $7~k$ modes). The ELU activation function is used at each hidden layer neurons and, {\it Adam optimizer} with learning rate $\eta=10^{-3}$ used as the gradient descent algorithm. The performance of the network was checked at each training epoch for both training and the validation set. The ANN reached $\approx 98\%$ validation accuracy within $<1000$ training epochs. Fig. \ref{fig:ann_powerspectrum_emulation} shows a comparison between the test set 21-cm PS (simulated) and corresponding the ANN emulations of the same. The emulator was able to achieve similar level of accuracy $\approx 98\%$ in the test prediction as well. Thereafter, the emulator was considered to be ready to be implemented in the Bayesian Inference pipeline. 
\footnotetext{\href{https://github.com/himmng/EmuPBk/tree/master/docs/source/test_emulations}{https://github.com/himmng/EmuPBk/}{\label{git2}}}

\subsection{Emulating 21-cm Bispectrum}
We take a similar approach in developing the ANN-based 21-cm bispectrum emulator. The bispectrum data sets were simulated for the same $550$ LH samples (see Fig. \ref{fig:parameter_space}). As discussed earlier, we consider bispectra $\Delta^3$ to be parameterized using triangle size and shapes, i.e. $(k_1,\,n,\,\cos{\theta})$. The training here uses the simulated bispectrum estimates at $5$ different sizes having $k_1$ values in the range $(0.19 - 1.50)\, {\rm Mpc}^{-1}$ and for every $k_1$, we consider $11$ and $10$ linearly-spaced bins for $n$ and $\cos{\theta}$ respectively (see \S\ref{2} for details). Thus, for each $k_1$ value, we have $110$ bispectrum estimates. However, In this work we only consider on the specific bisepctrum values that satisfy the unique triangle conditions (eq. \ref{eq:cond}). It restrict the total bispectrum values to $66$ for every $k_1$ mode. A visual  representation of unique triangle shapes in the $n-\cos{\theta}$ parameter space is demonstrated in the right panel of Fig.\ref{fig:unique_triangle_space}. Therefore the resulting size of the unique triangle bispectra becomes $5\times66$ (considering $5~k_1$ modes) for each set of reionization parameters. This training data is flattened and scaled before training the ANN.

\begin{figure*}
     \includegraphics[scale=0.55]{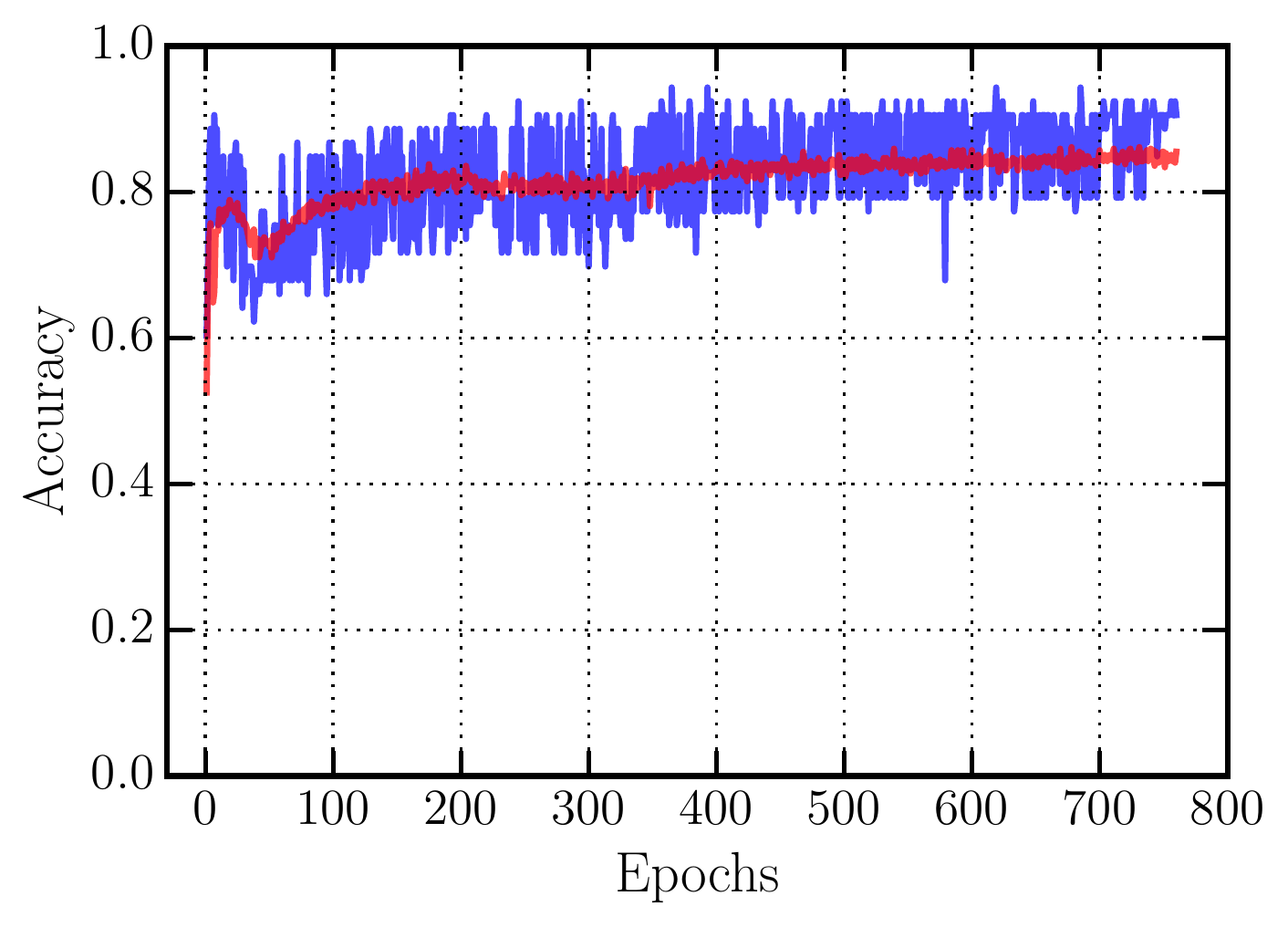}
     \includegraphics[scale=0.55]{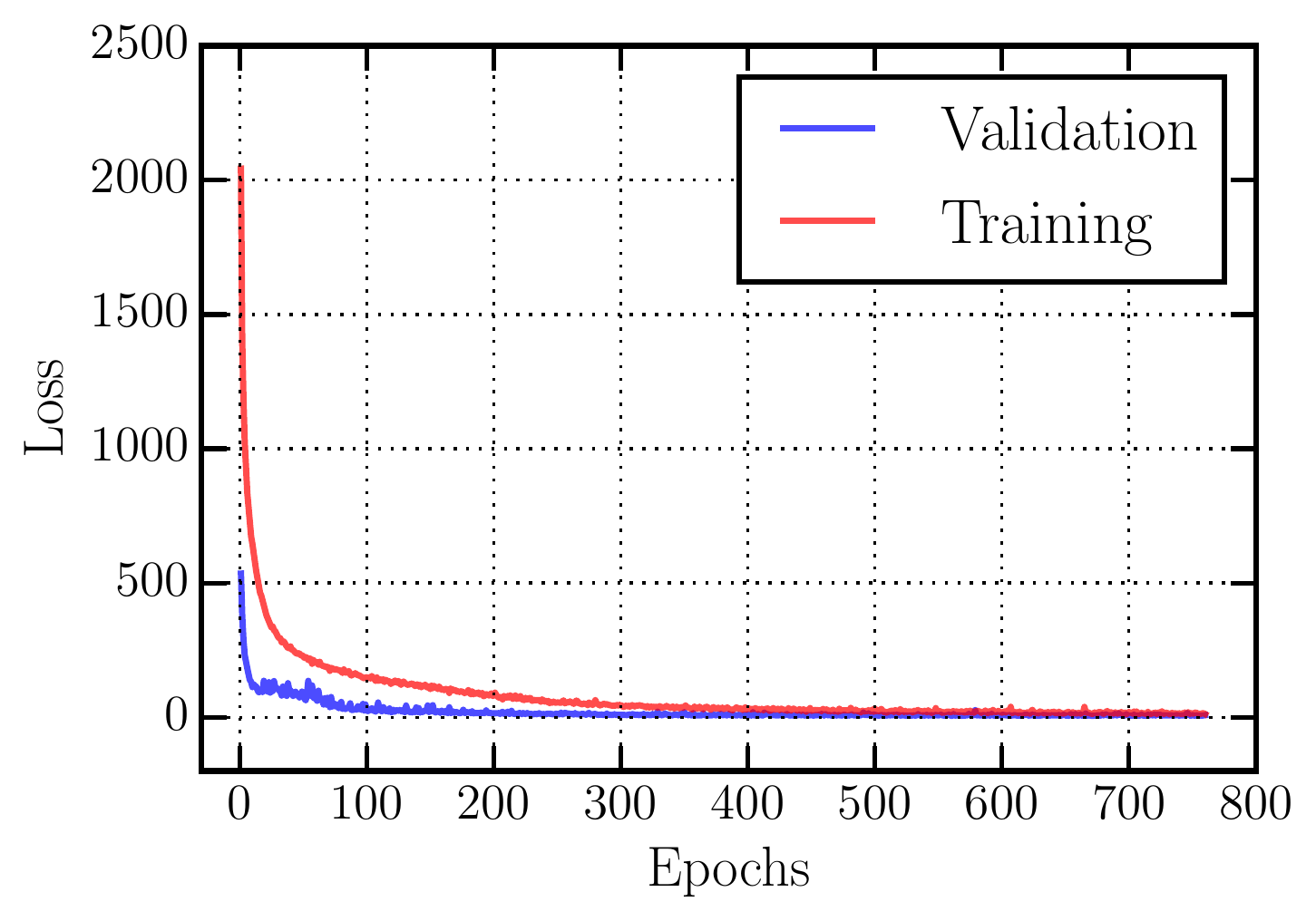}
     \caption{Shows the accuracy (left panel) and loss (right panel) of our ANN model (for bispectrum) as a function of training epochs. Here we show the results for training and validation sets both.}
     \label{fig:ann_accuracy_and_loss}
\end{figure*}

\begin{figure*}
\centering
    \includegraphics[scale=0.44]{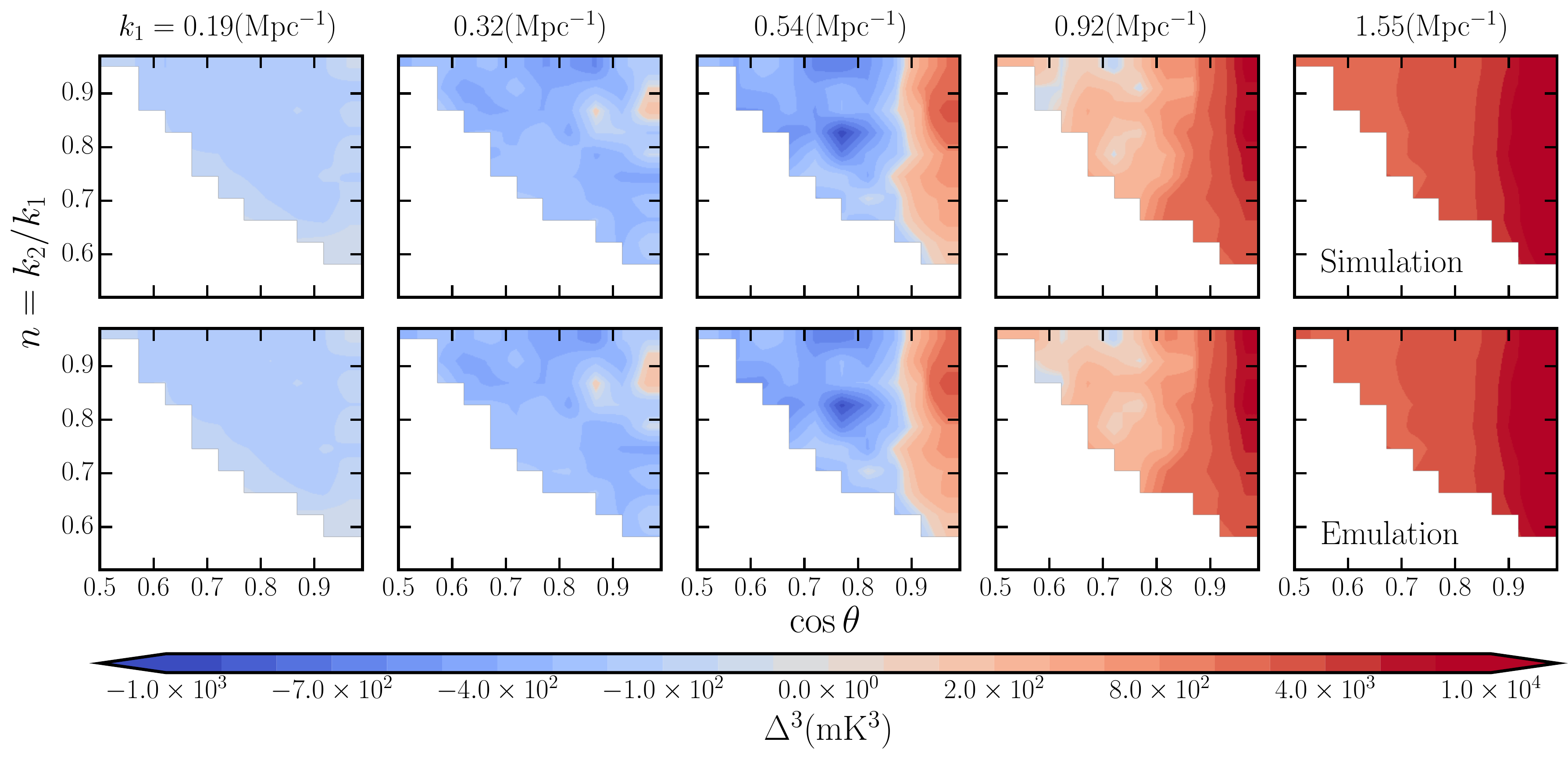}
    \caption{Shows a comparison between the simulated (top row) and emulated (bottom row) bispectrum for a test set model with $\Mmin = 35.0(\rm 10^9 M_\odot)$, $\Nion = 149.64$, $\Rmfp= 49.0$ $({\rm Mpc})$ and $\xb = 0.83$. We show the results for $5$ different sizes ($k_1$ values) of triangles in different columns. Similar plots for other test set models are available on our GitHub page\textsuperscript{\ref{git2}}.}
    \label{fig:ann_bispectrum_prediction}
\end{figure*}

Our bispectrum emulator consists of $2$ hidden layers with $66$ and $256$ nodes respectively. Here we have used the same ELU activation function at every nodes in the hidden layers and {\it Adam optimizer} with a learning rate $\eta=10^{-4}$ for gradient descent algorithm. The training, validation and test sets have the same reionization parameter sets as used in PS emulator. The performance of the neural network is demonstrated in Fig. \ref{fig:ann_accuracy_and_loss}, which shows the variation of MSE loss (right panel) and accuracy (left panel) for the training and validation set. The validation loss can be smaller than the training loss in scenarios where the validation set is relatively less featured than the training set. We see that both the training and validation sets crosses the $85\%$ accuracy mark, and for the test set it is $\approx 93.3\%$. The Fig. \ref{fig:ann_bispectrum_prediction} presents a visual comparison between the 21-cm BS predictions from our emulator (bottom panels, for $5$ different $k_1$ bins)  with one simulated 21-cm BS from the test set. We observe that our emulator predictions closely match with the simulated bispectrum, both qualitatively and quantitatively. This confirms that the performance of our bispectrum emulator is reasonably good.

\section{Inferring Reionization Parameters}\label{5}

\subsection{Bayesian Inference Framework}
The questions that we want address in this work are the following -- `How well can one constrain the reionization model parameters using the measured EoR 21-cm PS and BS? Does including all unique $k$-triangle shapes while estimating bispectra improves our statistical inferences over using power spectrum alone or bispectrum for a few specific $k$-triangle shapes?'. In order to answer these questions, we use a Bayesian inference approach to constrain the EoR model parameters and predict errors in the inferred parameter values. According the Bayes' theorem, we can express the posterior probability distribution of the parameters as (see e.g. \citep{Sharma_2017}):
\begin{equation}
     p(\boldsymbol{\alpha}|D,M) = \frac{p(D|\boldsymbol{\alpha},M)~\Pi(\boldsymbol{\alpha}|M)}{p(D|M)}~,
\label{eq:bayes}
\end{equation}
where $\boldsymbol{\alpha}$ is the parameter vector and $D$ and $M$ denote the data and the model, respectively. For a specific model, the evidence $p(D|M)$ becomes a constant normalization factor and the posterior solely depends on the product of the likelihood $L \equiv p(D|\boldsymbol{\alpha},M)$ and the prior distribution for $\boldsymbol{\alpha}$ i.e. $\Pi(\boldsymbol{\alpha}|M)$. Our analysis presumes a multivariate Gaussian likelihood for the data and the logarithm of which is expressed as:
\begin{equation}
    \ln{L} = -\frac{1}{2} [\vec{d}_{\rm ref}-\boldsymbol{\mu}]^{\rm T} [\cov]^{-1} [\vec{d}_{\rm ref}-\boldsymbol{\mu}] - \frac{1}{2}\ln{(2\pi \det{\cov})}~.
\label{eq:log-lik}
\end{equation}
Here $\vec{d}_{\rm ref}$ denotes reference data array, $\boldsymbol{\mu} \equiv \boldsymbol{\mu}(\boldsymbol{\alpha})$ denotes the model observable corresponding to a set of parameter values $\boldsymbol{\alpha}$ and $\cov$ is the error covariance associated with $\vec{d}_{\rm ref}$. 

Based on the above log-likelihood, we design a pipeline to identify the most likely EoR model parameter values using Markov Chain Monte-Carlo (MCMC) random walks in a vast parameter space. Our pipeline is written in \texttt{Python} and employs \texttt{CosmoHammer}\footnote{Publicly available at: \href{https://pypi.org/project/cosmoHammer/}{https://pypi.org/project/cosmoHammer}} \citep{Akeret_2013}, an Affine invariant MCMC ensemble sampler \citep{Goodman_2010}, to perform the model parameter estimation. To test the performance of the pipeline we input either the binned 21-cm power spectrum $\Delta^2(k_i)$ or the binned 21-cm bispectra $\Delta^3_i(k_1, n, \cos{\theta})$, computed from our EoR simulation (see \S\ref{3}), as the reference data $\vec{d}_{\rm ref}$. The respective error covariances $\cov$ are also estimated (we discuss it in the following subsection) and supplied to the pipeline. The MCMC sampler is then run with $20$ random walkers, each of which takes $20000$ steps as a standard sampling chain. Thus a single MCMC run samples $4\times 10^5$ points in the EoR parameter space for which it requests $\boldsymbol{\mu} \equiv \Delta^2(k)$ or $\Delta^3(k_1,n,\cos{\theta})$ from our signal emulators (see \S\ref{4}). Assuming an uniform prior, i.e. $\Pi(\alpha|M)= \rm constant$, within the parameter range used for signal emulation, we run our Bayesian inference pipeline separately for the PS and the BS data. Note that, we find each of the MCMC chain converges merely within $2000$ steps for BS. In case of the PS the steps required is even less $(\sim 100)$. Our inference pipeline make use of the power spectrum emulated at all of the seven $k$ bins as shown in Fig. \ref{fig:ann_powerspectrum_emulation}. On the other hand, the bispectrum-based pipeline utilize all unique shapes of $k$-triangles. However we consider triangles only with three different sizes $k_1 = (0.19, 0.32, 0.54)~\impc$ for which the error covariance is expected to be within the detectable limits of the future telescopes, e.g. the SKA. The bispectrum for triangle bins having larger $k_1$ modes are system noise dominated whereas the smaller $k_1$ bins suffer from sample variance (Fig. \ref{fig:noise_and_sample_variance}). 

\subsection{Power spectrum and Bispectrum error covariances}
The error in the observables is a major source of uncertainties in the inferred parameter values. Hence prior estimate of the errors in the observables (here the EoR 21-cm PS and BS) is required to predict the posterior of the parameters. In addition to the inevitable sample variance, the EoR 21-cm signal has contamination from the large foregrounds \citep{Ali_2008, Ghosh_2012}, system noise, calibration errors etc. In this analysis, we assume that the foregrounds, calibration errors etc. have been perfectly modelled and completely removed from the observed signal. Hence total observed signal $\Delta_t(\kk)$ is now a sum of the EoR 21-cm signal $\Delta_{\rm b}(\kk)$ and the Gaussian system noise $\Delta_{\rm N}(\kk)$. Therefore in this case, the total covariance $\cov_t$ will have contributions only from the sample variance and the system noise. Assuming the 21-cm signal and the system noise to be disjoint, we add the both of these error contributions in quadrature at the same bin to obtain total covariance
\begin{equation}
    \cov_t(i) = \cov_{\rm SV}(i) + \cov_{\rm N}(i)~,
    \label{eq:tot_cov}
\end{equation}
where $\cov_{\rm SV}(i)$ and $\cov_{\rm N}(i)$ denotes respectively the sample variance and the system noise contribution to the total error in the $i$-th bin (power spectrum or bispectrum). Note that our analysis assumes that the measurements at any two bins are mutually uncorrelated. This simplifies the error computation and results into a diagonal covariance matrix{\footnote{Note, we have assumed here that the foregrounds and other systematic have been perfectly modelled and removed from the observed data, which may not be the case in a realistic scenario. Presence of any residual foregrounds in the data in principle can introduce non-zero off-diagonal elements in the error covariance matrix (see e.g. \cite{watkinson21a}).}}. In our subsequent discussions, we therefore use `variance' instead of `covariance' for $\cov$ without any loss of generality.

Here we use a simplified form of sample variance $\cov_{\rm SV}(P_i)=[\Delta^2(k_i)]^2/\Nk$ for the bin averaged power spectrum $P(k_i)$. Although the trispectrum contribution to the sample variance is considerably large at the large $k$ bins \citep{Mondal_2015}, we ignore this term in our $\cov_{\rm SV}(P_i)$ for reducing the computational cost. Also dropping the trispectrum contribution would underpredict the total error estimates (after including system noise) by at most $\sim 10\%$ for the $k$ range and redshift considered here \citep{Shaw_2019}. Next, the system noise contribution to the power spectrum error covariance can be written as \citep{Scoccimarro_2004, Liguori2010}
\begin{equation}
    \cov_{\rm N}(P_i) = 2\frac{V_f}{V_P} [\Delta^2_{\rm N}(k_i)]^2~,
    \label{eq:PS_noise_cov}
\end{equation}
where $\Delta^2_{\rm N}(k_i)=k_i^3 P_{\rm N}(k_i)/(2\pi^2)$, $V_f = 2\pi^3/V$ and $V_P = 4\pi k_i^2 \Delta k_i $ with $\Delta k_i$ being the width of $i$-th bin. In this paper, we consider the system noise corresponding to $1000$ hrs of mock observations using the upcoming SKA-Low telescope. Any radio-interferometric observation measures signal at specific baselines and particular frequency channels. Considering any pair of antenna, the baselines are the projection of their separation on the sky plane in units of observing wavelength. The baselines at any particular redshift (or frequency) are directly proportional to the perpendicular to the LoS component, $\kk_\perp$, of the wave vector $\kk$ in the sky. Assuming the SKA-Low observing at $-30^\circ$ declination for $8$ hrs/night, we simulate the baseline distribution at $z=8$ using the current proposed configuration of SKA-Low with $512$ stations \citep{SKA_Low_v2}. Note that, here the stations are a collection of hundreds of dipoles which acts like an antenna having diameter $35~{\rm m}$. Following the prescription described in Section $3$ of \cite{Shaw_2019}, we uniformly grid the $\kk_\perp$ space with a baseline spacing equivalent to the station diameter. Next we collapse baselines onto their nearest grid points which have uncorrelated measurements. Assuming ergodicity along the LoS direction, we fill the whole $\kk$ space by putting the same gridded baselines at each grid along $k_\parallel$ axis. Note that, we only consider the chunk of the signal within a frequency bandwidth of $8~{\rm MHz}$ centered at $157.82~{\rm MHz}$ (corresponds to $z=8$) having channel width $0.1~{\rm MHz}$. We finally compute the system noise power spectrum at every grid-point in the reconstructed $\kk$ space following eq. (1) of \cite{Shaw_2019}. We do not go into the details of the noise power spectrum calculations here and readers are referred to the Section $3$ of \cite{Shaw_2019} for a more detailed description of the same. We finally bin the $\kk$ space into logarithmically spaced spherical shells to and compute the spherically averaged system noise power spectrum $P_{\rm N}(k_i)$ which we use in our analysis to compute the system noise error variance (eq. \ref{eq:PS_noise_cov}). Following eq. (\ref{eq:tot_cov}), we combine $\cov_{\rm SV}(P_i)$ and $\cov_{\rm N}(P_i)$ to obtain the total error variance in the power spectrum measurement. 

\begin{figure*}
    \includegraphics[scale=.58]{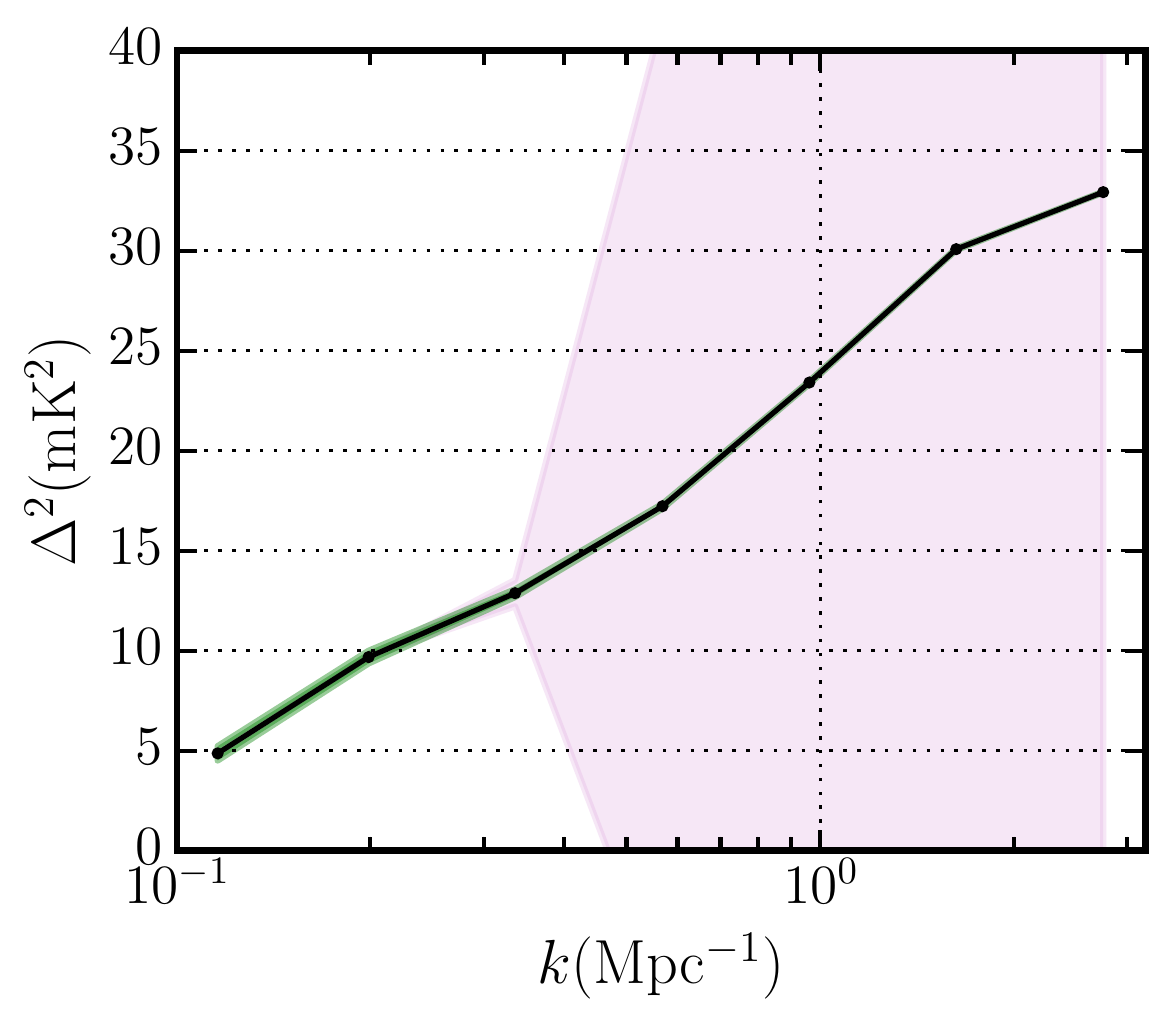}\qquad \quad
    \includegraphics[scale=.58]{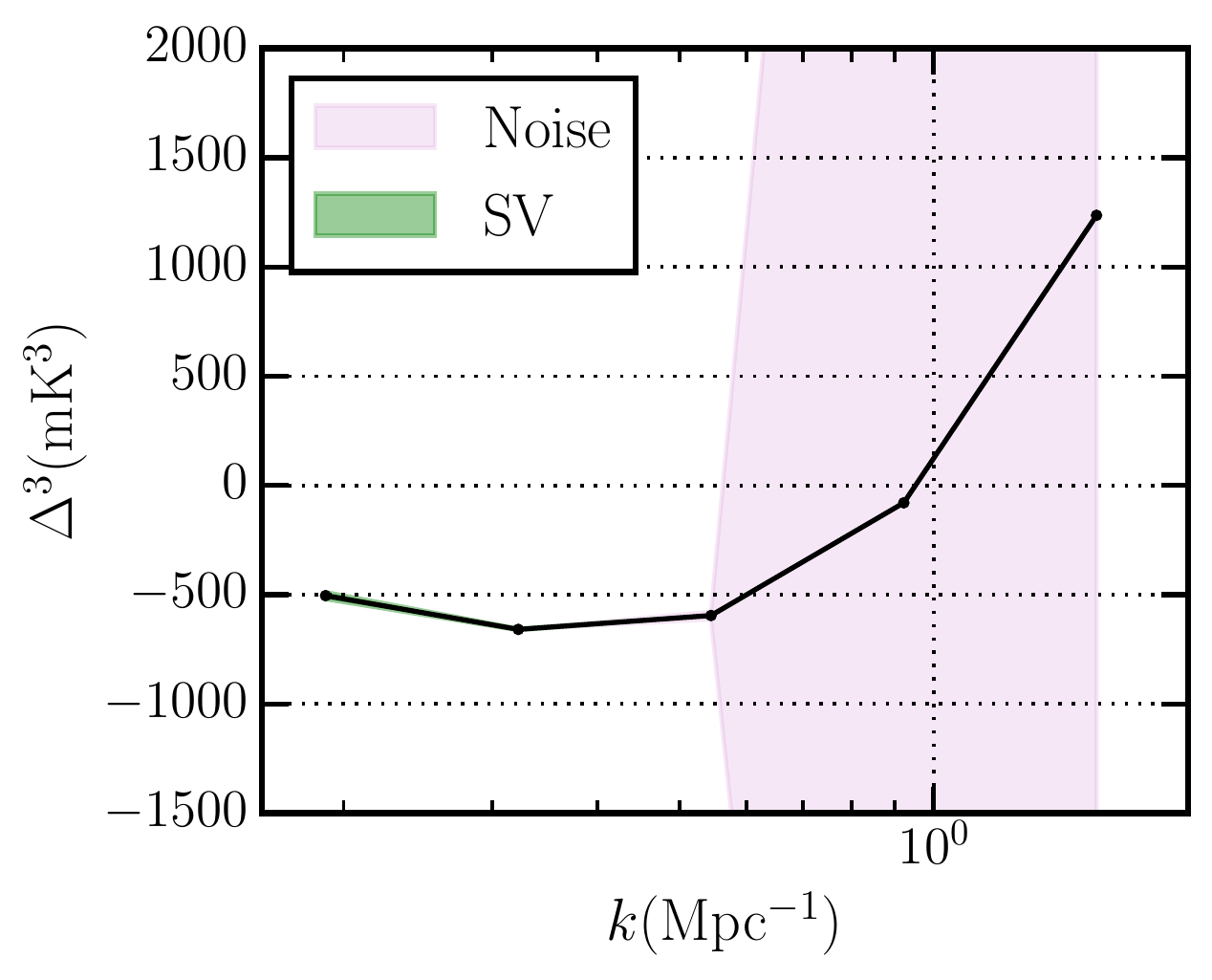}
    \caption{Shows the power spectrum $\Delta^2(k)$ (left panel) and the equilateral bispectrum $\Delta^3(k_1, n=1, \cos{\theta}=0.5)$ (right panel) by the solid black lines for a reionization model $(\Mmin, \Nion, \Rmfp) = [7.1(10^9{\rm M}_\odot), 75.18, 35.36(\rm Mpc)]$ having $\xb = 0.63$ at $z=8$. The green shaded regions (narrower) show the corresponding SV estimates whereas the purple shaded regions show the system noise contribution to the error variance corresponding to $1000$ hrs of SKA-low observations.}
    \label{fig:noise_and_sample_variance}
\end{figure*}

The left panel of Fig. \ref{fig:noise_and_sample_variance} shows the power spectrum corresponding to a model \\
$(\Mmin, \Nion, \Rmfp) = (7.1\times 10^9~{\rm M}_\odot, 75.18, 35.36)$ for which $\xb = 0.63$ at $z=8$. We also show the $\cov_{\rm SV}$ as green shaded region and the $\cov_{\rm N}$ as the purple shaded region around the signal power spectrum. Note that the system noise contribution to the power spectrum has been estimated for the $1000$ hrs of future SKA-low observations. We note that the EoR 21-cm PS is detectable for $k < 0.4~\impc$ where the SV and system noise contributions are much below the signal.  The system noise contribution to the error suddenly increases for $k\gtrsim 0.4 ~\impc$ leaving the power spectrum undetected at those scales.

We next discuss the error computations of the bispectrum used in our analysis. Drawing the analogy from the power spectrum sample variance, we can approximately write the sample variance of the bin-averaged bispectrum $B_i(k_1,n, \cos{\theta})$ as $\cov_{\rm SV}(B_i)= [\Delta^3_i(k_1,n,\cos{\theta})]^2/N_{\rm tri}$. Here binning is done in shape $(n,\cos{\theta})$ and size $(k_1)$ parameter-space of triangles, and $N_{\rm tri}$ is the number of triangle in the any $i$-th triangle bin (eq. \ref{eq:binbs}). Note that, here we choose the a simplistic form of $\cov_{\rm SV}(B_i)$ and ignore other higher-order terms in order to reduce the computational cost. It is advisable to use the complete sample variance \citep{Mondal_2021} wherever high precision is mandatory, and we defer this to our future works. However we again emphasise that the sole motivation of this analysis is to demonstrate the fact that if  bispectrum estimates for all unique triangle shapes are used a much better constraints on the reionization parameters can be obtained compared to the scenario when  only power spectrum is used, and therefore an order of magnitude estimation of errors would be sufficient for our purpose. Next, we compute the system noise contribution to the bispectrum error variance which can be written as \citep{Scoccimarro_2004, Liguori2010}
\begin{equation}
    \cov_{\rm N}(B_i) = s_B~\frac{V_f}{V_B}~\Delta^2_{\rm N}(k_1) \Delta^2_{\rm N}(k_2) \Delta^2_{\rm N}(k_3)~,
    \label{eq:BS_noise_cov}
\end{equation}
where $s_B = 1$ for general triangles and $V_B \approx 8\pi^2 (k_1 \Delta k_1) (k_2 \Delta k_2) (k_3 \Delta k_3)$ with $(k_1,k_2,k_3)$ is the mean side-lengths of the triangles within the $i$-th bin with $(\Delta k_1,\Delta k_2,\Delta k_3)$ being the corresponding extent of the bin in $\kk$ space. Note that, given the mean shape $(n,\cos{\theta})$ of a triangle bin we can easily compute back the $k_2$ and $k_3$ using the triangle geometry. In order to estimate $\cov_{\rm N}(B_i)$ (eq. \ref{eq:BS_noise_cov}), we obtain $\Delta^2_{\rm N}(k_a)$ by interpolating $\Delta^2_{\rm N}(k_i)$ which has been discussed earlier. Finally we add the errors into the quadrature to obtain the total error variance in the binned bispectrum. 
The right panel of Fig. \ref{fig:noise_and_sample_variance} shows the equilateral bispectrum corresponding to aforementioned model where $\xb = 0.63$. The corresponding $\cov_{\rm SV}$ and the $\cov_{\rm N}$ are also shown respectively by the green and the purple shaded regions around the signal bispectrum curve. Here we particularly choose the equilateral triangles in order to demonstrate the size dependence of the bispectrum. The equilateral triangle is expected to have largest system noise contribution to the error among all possible shapes. We find that the equilateral bispectrum can be detected for triangles having sizes $k_1 < 0.6~\impc$ where the SV and system noise contributions are negligibly small as compared to the signal. However the system noise contribution to the error drastically increases for $k\geq 0.6 ~\impc$ making the bispectrum difficult to detect at those scales.

\subsection{EoR Parameter estimation}
\begin{table*}
    \centering
    \begin{tabular}{ccccc}
        \hline
            & $\Mmin~(10^9{\rm M}_\odot)$ & $\Nion$ & $\Rmfp~({\rm Mpc})$ & $\xb$ \\
            \hline
            \hline
        Set-1 & $35.0$ & $149.64$ & $49.00$ & $ 0.83$  \\ 
        Set-2 & $7.1$ & $75.18$ & $35.36$ & $ 0.63$  \\
        Set-3 & $15.2$ & $45.45$ & $32.09$ & $ 0.88$  \\
        \hline
    \end{tabular}
    \caption{Lists the fiducial values of the EoR parameter sets for the three different models considered here along with their average neutral fraction $\xb$. }
    \label{tab:models}
\end{table*}

In this subsection we present the constraints over the EoR model parameters that we obtain from our MCMC analysis. Note that, we consider bispectrum for all unique shapes of triangles in this analysis which allows us to exploit the maximum information contained within the bispectrum. We compare our results against the constraints obtained using only the power spectrum statistics. In addition to it, we also show results considering the bispectrum only for isosceles triangles which is expected to retain important features of the non-Gaussianity present in the EoR 21-cm signal \citep{Majumdar_2018, Majumdar20, hutter19, watkinson_2021}. Note that here we consider only the L-isosceles triangles which encompasses all the shapes between the equilateral and squeezed triangles. In this analysis, we consider three different input observed 21-cm signal scenarios corresponding to EoR parameter values and average neutral fraction $\xb$ as listed in Table \ref{tab:models}. 

\begin{figure}
    \centering
    \includegraphics[scale=0.49]{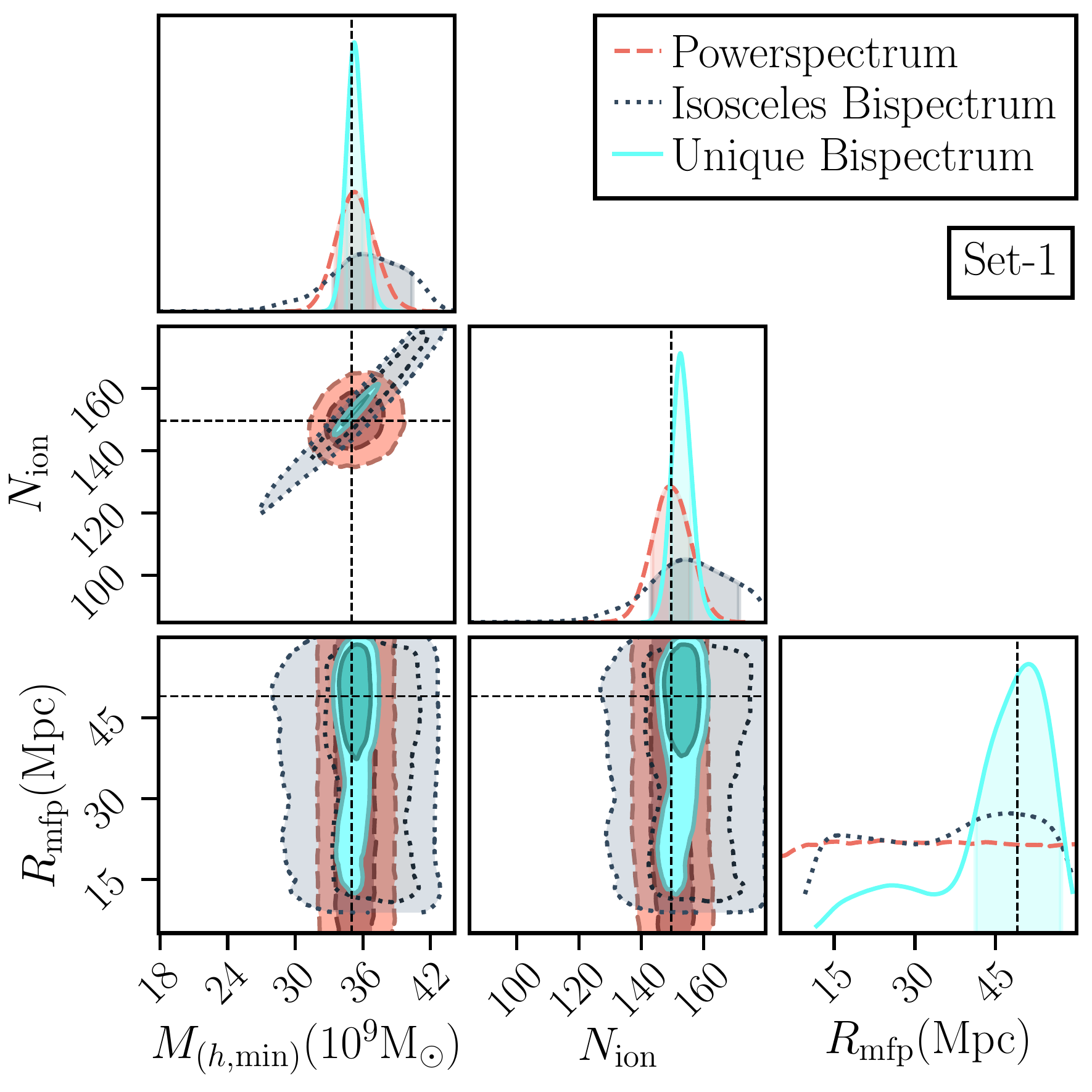}
    \caption{Shows the posterior of the model parameters obtained using the power spectrum (red), the isosceles bispectrum (gray) and all unique bispectrum (cyan). The regions bounded by the inner and the outer contours represent $1\sigma$ and $2\sigma$ levels respectively. The results here corresponds to a model with true parameter values $(\Mmin,\Nion,\Rmfp) = (35 \times 10^9\, {\rm M}_\odot ,149.64, 49\,{\rm Mpc})$ having $\xb = 0.83$ (see Table \ref{tab:models}).}
    \label{fig:joint_estimation_figure_1}
\end{figure}

Fig. \ref{fig:joint_estimation_figure_1} shows the marginalized likelihood plots for the three EoR parameters (see \S\ref{3}) corresponding to Set-1 (see Table \ref{tab:models}). We show $1\sigma$ and $2\sigma$ contours of the joint-likelihood for every pair of the parameters in the off-diagonal panels. However the posterior of individual parameters are shown in the diagonal panels. The dashed (vertical and horizontal) lines correspond to the true value of the  parameters. Considering $\Mmin$ and $\Nion$ first, we find a positive correlation between them. This is expected as these two parameters have direct but inverse impact on how fast the reionization progresses. However we do not notice any correlations between $\Rmfp$ and other two parameters. We find that using bispectrum of all unique shapes of triangle provide substantially tighter constraints on $\Mmin$ and $\Nion$ as compared to the constraints obtained using only the power spectrum. The percentage uncertainties in $\Mmin$ and $\Nion$ with respect to their true values are respectively $10\%$ and $8.2\%$ when constrained using power spectrum (see Table \ref{tab:constraints}). These uncertainties drop down to $4.5\%$ and $4.3\%$ respectively when constrained using unique bispectrum. Considering $\Rmfp$, we see that the unique bispectrum is able to recover the parameter within $32.4\%$ uncertainty whereas the power spectrum fails to provide a constraint. This is simply because of the fact that reionization is not very sensitive to $\Rmfp$ \citep{Schmit_2018, Binnie_2019, Shaw_2020} and the bispectrum is wealthier in information than the power spectrum. 

\begin{table*}
    \centering
    \begin{tabular}{ccccc}
        \hline
                & Statistical Estimator & $M_{(h,\rm min)}(10^9 ~\rm M_\odot)$ & $N_{\rm ion}$ & $R_{\rm mfp}(\rm Mpc)$ \\
                \hline \\
                & Power spectrum & $35.2^{+1.7}_{-1.7}$ & $149.2^{+6.6}_{-5.7}$ & $-$ \\ \\
            Set-1    & Isosceles Bispectrum & $36.0^{+4.4}_{-2.6}$ & $154^{+17}_{-11}$ & $-$ \\ \\
                & Unique Bispectrum & $35.23^{+0.80}_{-0.76}$ & $152.6^{+3.5}_{-3.0}$ & $51.1^{+6.0}_{-9.9}$ \\ \\
                \hline \\
               & Power spectrum & $7.19^{+0.73}_{-0.67}$ & $76.4^{+5.7}_{-5.6}$ & $-$ \\ \\
        Set-2    &    Isosceles Bispectrum & $8.2^{+2.1}_{-1.8}$ & $81^{+11}_{-10}$ & $-$ \\ \\
             &   Unique Bispectrum & $7.29^{+0.45}_{-0.43}$ & $76.5^{+2.6}_{-2.8}$ & $35.6^{+3.1}_{-2.8}$ \\ \\
                \hline \\
            &    Power spectrum & $15.8^{+1.3}_{-1.6}$ & $46.8^{+6.6}_{-6.1}$ & $-$ \\ \\
        Set-3    &    Isosceles Bispectrum & $16.4^{+1.9}_{-2.1}$ & $46.2^{+5.2}_{-3.1}$ & $35^{+12}_{-11}$ \\ \\
            &    Unique Bispectrum & $15.42^{+0.37}_{-0.42}$ & $46.69^{+0.92}_{-0.79}$ & $33.6^{+2.2}_{-3.6}$ \\ \\ 
                \hline
    \end{tabular}
    \caption{Lists the inferred EoR model parameters with $1 \sigma$ errors estimated using MCMC. The true values of the parameters are provided in the Table \ref{tab:models}.}
    \label{tab:constraints}
\end{table*}

Considering next the isosceles triangle bispectrum, we note that the uncertainties in the parameters increases even when compared to the power spectrum constraints. Here the percentage uncertainties for $\Mmin$ and $\Nion$ are respectively $20\%$ and $18.7\%$ which is roughly double what we have for the power spectrum (see Table \ref{tab:constraints}). Isosceles bispectrum is also unable to constrain $\Rmfp$ just like the power spectrum. We also note that the deviation between the recovered and the actual values are also larger than those when we use the power spectrum or the unique bispectrum. This behaviour qualitatively agrees with the findings of a recent work by \cite{watkinson_2021} where the authors have only considered the isosceles triangles to estimate the reionization parameters. The bispectrum, being a three-point statistics, have larger error in its estimates than the power spectrum. Additionally, the constraining power of an estimator is simultaneously crucial for the parameter estimation. Although the isosceles triangles mostly cover the shape dependence of the bispectrum, they may not be sensitive enough to the EoR model parameters. This is a plausible explanation of the poor performance of the isosceles triangles. However it is possible that bispectrum at some other triangle shapes are highly efficient in terms of sensitivity over parameters and the noise. Using such shapes can provide improved constraints over the parameters. However finding such triangle shapes is non-trivial, and we generally expect such efficient shapes to vary from model to model (of the signal) and also across the different stages of reionization. This will be interesting to investigate and we defer it to our future studies. Inclusion of all unique shapes of $k$-triangles in our analysis ensures that we always take into account those shapes which have higher sensitivity to the EoR parameters.

\begin{figure*}
     \includegraphics[scale=0.4]{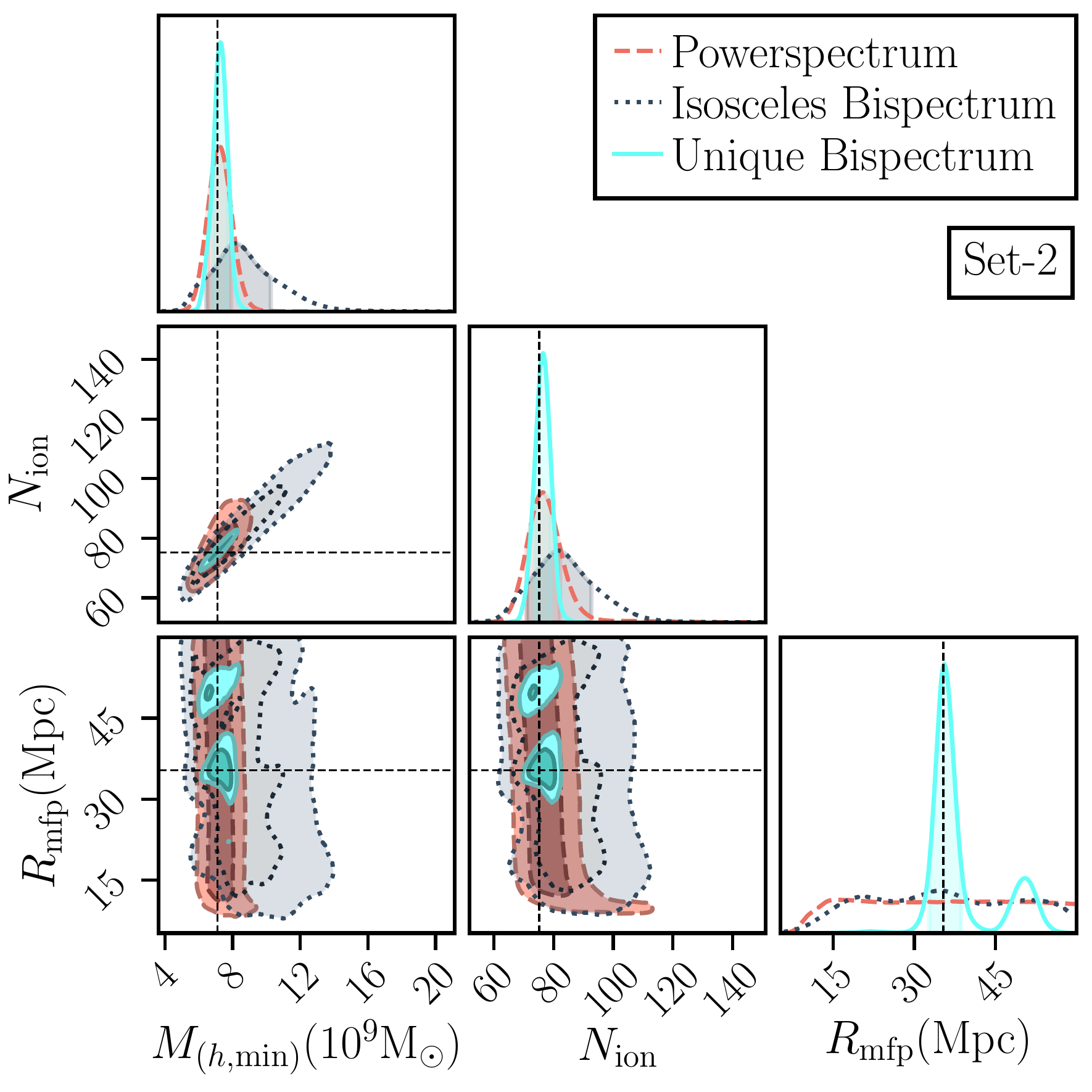}
     \qquad \quad
     \includegraphics[scale=0.4]{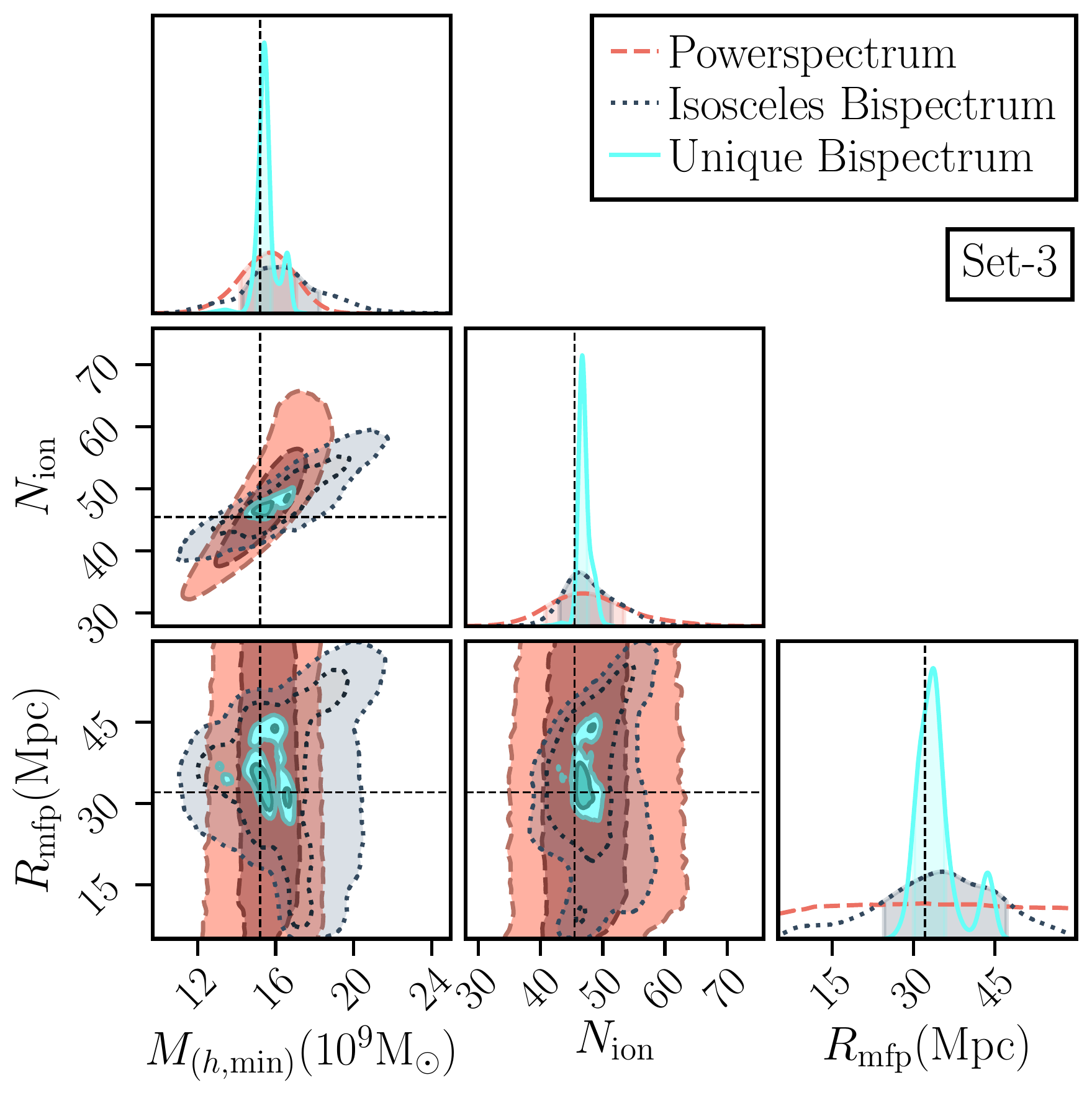}
     
     \caption{Shows the posteriors of parameters as in Fig.
     \ref{fig:joint_estimation_figure_1}.
     The contours show $1\sigma$ and $2\sigma$ confidence contours. The left and the right plots corresponds to the reionization models with true parameter values $(\Mmin(\rm 10^9M_\odot), \Nion,  \Rmfp(\rm Mpc); \xb) = (7.1, 75.18, 35.36; 0.63)$ (Set-2) and $(15.2, 45.45, 32.09; 0.88)$ (Set-3) from Table \ref{tab:models}.}
     \label{fig:joint_estimation_figure_2}
\end{figure*}

Next for validating our arguments, we show similar likelihood plots in Fig. \ref{fig:joint_estimation_figure_2} for two other input signal parameter sets, namely Set-2 and Set-3 (see Table \ref{tab:models}). Here Set-2 (left plot) corresponds to an early reionization scenario with respect to Set-1, whereas Set-3 (right plot) represents a late reionization scenario. Considering the three signal statistics, we find that their comparative behaviour remain qualitatively similar for these two sets as well. We also list the recovered parameter values and the corresponding uncertainties in Table \ref{tab:constraints}. 

Note that, here we present results only for those EoR models which have $\xb > 0.5$ as the parameter constraints are not very encouraging for the models with $\xb \lesssim 0.5$. This is simply because the emulator exhibits a poor performance for models with $\xb \lesssim 0.5$ as we have access to only a few models $(\approx 15)$ to train the emulator in the lower $\xb$ domain (see Fig. \ref{fig:parameter_space}). This is a consequence of choosing a regular grid spacing in the Latin-hypercube sampling of the parameters to build the training sets for our signal emulators. The reionization history has a complicated non-linear dependence over the parameters which yield the training sets having non-uniform sampling of $\xb$. Our training sets here has more points with larger neutral fraction ($\xb>0.5$). It is therefore essential that an efficient gridding scheme should be chosen for the parameter space to build a more comprehensive training set for the signal emulators. We plan to develop and incorporate such a strategy in future follow-up work.

In conclusion, the bispectrum with all unique shapes provide a better constraint on the EoR parameters as compared to the power spectrum. We expect a further improvement in inferred parameter values if one combines power spectrum and all unique bispectrum measurements and performs a joint parameter estimation.


\section{Summary and Discussion}\label{7}

The EoR 21-cm signal is a highly non-Gaussian field, and this non-Gaussianity holds the potential to provide a deeper understanding of the EoR. The bispectrum is the lowest order statistic which can quantify the time evolving non-Gaussianity in the EoR 21-cm signal. The computation of bispectrum $B(\kk_1,\kk_2,\kk_3)$ involves measurements of the signal at three wave vectors which by construction has to form a closed triangle (see eq. \ref{eq:binbs}). Therefore one can use the sizes $(k_1)$ and shapes $(n, \cos{\theta})$ of the triangles to parameterize the bispectrum (see eq. \ref{eq:shape}). Following \cite{Bharadwaj_2020}, we only consider triangles with unique shapes (eq. \ref{eq:cond}) in our analysis of the EoR 21-cm signal bispectrum. The aim of this paper is to demonstrate that using 21-cm bispectrum of all unique shapes one can provide a better constraint over the reionization model parameters than using just the power spectrum of the signal. Here we use the bin-averaged power spectrum (eq. \ref{eq:PS}) and bispectrum (eq. \ref{eq:binbs}) as our observable statistics of the EoR 21-cm signal to address this issue.

The work presented here uses Bayesian MCMC analysis in order to trace out both the joint and individual posterior probability distributions of the reionization model parameters for any given observed EoR 21-cm signal. Here we consider a semi-numerical \textit{inside-out} reionization simulation \texttt{ReionYuga} as our model for the EoR 21-cm signal, with three model (as well as simulation) parameters namely (1) $\Mmin$, the minimum halo mass that can host an ionizing source, (2) $\Nion$, the effective ionizing efficiency of the sources and (3) $\Rmfp$, the mean free path of the ionizing photons in the IGM. A detailed description of the parameters is presented in \S\ref{3}. Using this EoR simulation we further develop ANN-based emulators of the EoR 21-cm power spectrum and bispectrum. The emulators are used as a replacement of the simulated signal statistics in our MCMC based Bayesian parameter estimation pipeline. The ANN based emulators substantially enhances the speed of parameter estimation as compared to using the actual signal simulations. In this work, we use a total of $550$ sets (randomly sampled from the parameter space) of simulated signal of reionization to train (using $535$ set) and test (using $15$ sets) the signal statistic emulators. Our emulators exhibit remarkable reproducibility (Fig. \ref{fig:ann_accuracy_and_loss}) of the signal statistics on the test data sets with values very close to that of simulated statistics (Figs. \ref{fig:ann_powerspectrum_emulation} and \ref{fig:ann_bispectrum_prediction}). 

We finally use our pipeline to infer the EoR parameters for the three different sets (defined by combination of input parameter values) of  signals namely Set-1 $(\xb=0.83)$, Set-2 $(\xb=0.63)$ and Set-3 $(\xb=0.88)$, as listed in Table \ref{tab:models}. We consider both the sample variance and the system noise at $z=8$ corresponding to $1000$ hrs of future SKA-Low observations contributes to the total error covariance of the binned power spectrum and the binned bispectrum estimators. Our system noise computations incorporate the SKA-low baseline distribution corresponding to observing a patch of the sky at ${\rm DEC}=-30^\circ$ for $8~{\rm hrs/night}$.

The bispectrum measurements, spanning all unique shapes of triangles, is expected to capture vital astrophysical information inherent to the EoR 21-cm signal which may be very sensitive to any variation in the EoR parameters. This allows the uncertainties in the inferred parameter values to  be considerably smaller when inferred using all unique shapes of $k$-triangles compared to when inferred using only the power spectrum. This is clearly demonstrated by the posterior plots of the parameters obtained here and shown in Figs. \ref{fig:joint_estimation_figure_1} and \ref{fig:joint_estimation_figure_2}. The constraints over $\Mmin$ and $\Nion$ noticeably improve for all three parameter sets considered here (see Table \ref{tab:models}). For Set-1 $(\xb=0.83)$, using unique bispectrum estimates instead of power spectrum, reduces the relative uncertainties in $\Mmin$ and $\Nion$ roughly by a factor of $2$ (Table \ref{tab:constraints}). This reduction factor for the relative uncertainties becomes more than $4$ towards smaller neutral fraction (Set-3; $\xb=0.63$) where the non-Gaussianity in the field is expected to be larger. Also the most-likely parameter values, inferred using unique bispectrum measurements, agrees well within the $1 \sigma$ uncertainty limits (Table \ref{tab:constraints}). On the other side, the weak dependence of the EoR 21-cm signal on $\Rmfp$ makes it unconstrainable using power spectrum measurements during the early stages of reionization \citep{Shaw_2020}. Yet $\Rmfp$ can still have an imprint on the shapes of the ionized and neutral regions, however small. The bispectra at different unique shapes of triangles are sensitive enough to capture the variation due to $\Rmfp$. Therefore using all unique bispectrum allows us to constrain the $\Rmfp$ with a reasonable uncertainty (see Figs. \ref{fig:joint_estimation_figure_1} and \ref{fig:joint_estimation_figure_2}).

Despite of having excess information content, the bispectrum measurements of individual $k$-triangle shapes comes with a large statistical error than the power spectrum measurements. Also it is expected that the sensitivity of bispectrum to the model parameters to vary across the various $k$-triangle shapes $(n,\cos{\theta})$. There can be a few particular shapes which, owing to a strong parameter dependence, will be able to provide better constraints by overcoming large statistical errors. In general, such highly-sensitive shapes will be different for different reionization models and may also vary depending on the stage of reionization. Therefore one may not obtain a better result while inferring parameters using bispectra for only a few specific $k$-triangle shapes. We demonstrated this fact by performing parameter estimation using only L-isosceles $k$-triangle bispectra (Figs. \ref{fig:joint_estimation_figure_1} and \ref{fig:joint_estimation_figure_2}). We note that the constraints obtained are worse than the power spectrum only scenario. The uncertainties in individual parameters are also larger (Table \ref{tab:constraints}) as well as the most-likely values are significantly deviated from the true input values. Our results obtained for L-isosceles $k$-triangle bispectra only are consistent with the inference of \cite{watkinson_2021}. In principle, one can search for such highly-sensitive $k$-triangle shapes in order to make better predictions using bispectrum for limited triangle shapes. However this is a rather non-trivial task and will be very much model dependent as discussed earlier. Utilizing the bispectrum estimates for all unique triangle shapes for parameter estimation is therefore a rather simple and effective workaround to include maximum information about the signal in our parameter estimation exercise. This is where our analysis differs majorly with the previous work of \cite{watkinson_2021} who have restricted their analysis to the isosceles and equilateral triangles only. Besides, the bispectrum used in \cite{watkinson_2021} is normalized with the power spectrum whereas we do not use such normalization of bispectrum values. This allows us to exploit the large dynamic range of the bispectrum amplitude in this work. It would be interesting to validate our findings using more complicated EoR models in future. Please note that the analysis presented here is done on a fixed redshift. Including the signal statistics measured from multiple redshifts or observed 21-cm signal frequencies is expected to improve the constraints on the inferred EoR parameters. It is important to note that the results presented here do not consider effects of foreground contamination. However the foregrounds are inherent to the real observations. Ideally the foreground contamination are localized within a wedge-shaped region in $(k_{\perp}, k_{\parallel})$ plane (e.g. \cite{Datta_2010, Parsons_2012b}) and the area of the foreground wedge increases with redshift. Avoiding these foreground contaminated $k$ modes from bispectrum computation will reduce the number of triangle samples. This reduction in the triangles will be different for the bins of different shapes and sizes. For a particular shape bin the reduction in the triangle samples would be more for triangles of smaller size as the foregrounds are more dominant at small $k$ modes. Similarly, for a particular size, the reduction in triangle samples due to foreground avoidance will be the largest for bins near the squeezed limit ($\cos{\theta} \rightarrow 1,~n=1$) and smallest for the bins near equilateral limit($\cos{\theta}= 0.5,~n=1$). Therefore the foreground avoidance will increase the sample variance of the bispectrum which in turn increases errors in the inferred parameters. However foreground avoidance will parallely affect the power spectrum sample variance and thereby the corresponding errors in the parameters (see e.g. \cite{Shaw_2020}). A detailed comparative study of the effect of foregrounds on parameter estimation using power spectrum and bispectrum statistics would be an interesting work. Additionally, in order to make the predictions more realistic, one needs to incorporate foregrounds and other observational effects  (see e.g. \cite{watkinson21a, trott19}) along with a proper treatment of sample variance which affects the error estimates of the bispectrum. We plan to address many of these issues in our future follow up work.

\acknowledgments
All of the computations and simulations for this work were done using the computing resources available to the Cosmology with Statistical Inference (CSI) research group at IIT Indore. The authors would like to thank Rajesh Mondal for several fruitful discussions in line with this work. AKS would like to thank Raghunath Ghara for useful discussions. HT would like to acknowledge Prof. D. P. Agrawal and Prashant Agrawal for providing financial assistance during the project.

\bibliographystyle{JHEP_new}
\bibliography{bib}

\end{document}